\documentclass[aps,preprintnumbers,prd,twocolumn,superscriptaddress,nofootinbib]{revtex4} 

\usepackage{slashed}
\usepackage{graphicx}
\usepackage{color}
\usepackage{amsfonts}
\usepackage{subfigure}
\usepackage{pdflscape}
\usepackage{afterpage}
\usepackage{amsmath}
\usepackage{mathrsfs}
\usepackage[utf8]{inputenc}

\newcommand{\ie}{{\it i.e., }}
\newcommand{\eg}{{\it e.g., }}

\newcommand{\be}{\begin{equation}}
\newcommand{\ee}{\end{equation}}
\newcommand{\bea}{\begin{eqnarray}}
\newcommand{\eea}{\end{eqnarray}}

\newcommand{\ds}{\displaystyle}

\newcommand{\beq}{\begin{equation}}
\newcommand{\eeq}{\end{equation}}

\begin{document}

\preprint{ACFI-T17-20}

\title{Color Breaking Baryogenesis}
\author{Michael J. Ramsey-Musolf}
\email{mjrm@physics.umass.edu}
\affiliation{Physics Department, University of Massachusetts Amherst, Amherst, MA 01003, USA}
\affiliation{Kellogg Radiation Laboratory, California Institute of Technology, Pasadena, CA 91125, USA}
\author{Graham White}
\email{graham.white@monash.edu}
\affiliation{ARC Centre of Excellence for Particle Physics, Monash University, Victoria 3800, Australia}
\affiliation{TRIUMF, 4004 Wesbrook Mall, Vancouver, British Columbia V6T 2A3, Canada}
\author{Peter Winslow}
\email{pwinslow@physics.umass.edu}
\affiliation{Physics Department, University of Massachusetts Amherst, Amherst, MA 01003, USA}

\begin{abstract}
We propose a scenario that generates the observed baryon asymmetry of the Universe through a 
multi--step phase transition in which SU(3) color symmetry is first broken and then restored. 
 %Two mechanisms produce a baryon asymmetry: one is analogous to conventional electroweak baryogenesis; 
%the second involves spontaneous violation of $B-L$ conservation near the phase boundary.  
%The contribution from the electroweak mechanism dominates, while the spontaneous violation of 
%$B-L$ conservation leaves a small relic charge asymmetry that is several orders of magnitude below current observational bounds. 
A spontaneous violation of $B-L$ conservation leads to a contribution to the baryon asymmetry that becomes negligible in the final phase. The baryon asymmetry is therefore produced exclusively through the electroweak mechanism in the intermediate phase.
We illustrate this scenario with a simple model that reproduces the observed baryon asymmetry.  
We discuss how future electric dipole moment and collider searches may probe this scenario, though
future EDM searches would require an improved sensitivity of several orders of magnitude.
\end{abstract}


\maketitle

%%%%%%%%%%%%%%%%%%%%%%%%%%%%%%%%%%%%%%%%%%%%%%%%%%%%%%%%%%%%%%%%%%
\section{Introduction}
%%%%%%%%%%%%%%%%%%%%%%%%%%%%%%%%%%%%%%%%%%%%%%%%%%%%%%%%%%%%%%%%%%

%Following the discovery of a fundamental scalar boson at the LHC, the detailed exploration of the electroweak symmetry-breaking sector is now in full swing. Since its initial detection, subsequent investigation has determined that the properties and interactions of this new particle are highly consistent with those expected of the Standard Model (SM) Higgs boson. However, despite the strong agreement to the data, it is well known that the SM cannot yet be deemed complete. Perhaps the 

The origin of the cosmic matter-antimatter asymmetry remains one of the outstanding open questions at the interface of 
cosmology  with particle and nuclear physics. The Planck experiment determines that the baryon asymmetry of the 
Universe (BAU) is ~\cite{Ade:2013zuv} \begin{equation} 
\frac{n_B}{s} \equiv Y_B = (8.59 \pm 0.11) \times 10^{-11} 
\end{equation} 
where $n_B$ ($s$) is the baryon number (entropy) density. To dynamically generate the BAU one must fulfil three Sakharov conditions\cite{Sakharov:1967dj}: baryon number violation, C and CP violation, and a departure from equilibrium. The Standard Model (SM) cannot explain the matter-antimatter asymmetry as it fails to provide sufficient CP violation \cite{Gavela:1993ts,Huet:1994jb,Gavela:1994dt} and the required out-of-equilibrium conditions\cite{Gurtler:1997hr,Laine:1998jb,Csikor:1998eu,Aoki:1999fi}. As such many beyond SM scenarios have arisen to accommodate this need.

For many years, electroweak baryogenesis (EWBG) has been one of the most attractive scenarios for explaining the BAU \cite{Trodden:1998ym,Morrissey:2012db,White:2016nbo}. The main reason for this interest has been its testable nature due to its strong connection with the weak scale. However, successful electroweak baryogenesis requires new bosonic states with masses near the weak scale and significant couplings to the Higgs boson in order to generate a strongly first-order electroweak phase transition (EWPT). 
%Even at this stage in the LHC experimental program, many popular models (though not all) incorporating these features have already been shown to be in conflict with current measurements,
One of the most widely-considered possibilities, the minimal supersymmetric Standard Model (MSSM) with relatively light top squarks, appears to be in considerable tension with LHC data
 \eg see~\cite{Curtin:2012aa,Katz:2015uja} (however, see also \cite{Liebler:2015ddv}). In this context, it is worth asking if there are well motivated and testable modifications to the EWBG paradigm. 

The rich landscape of phase structures in condensed matter systems suggests that the thermal history of symmetries in the Universe might be more exotic than the conventional scenario involving a single instance of electroweak symmetry-breaking at a temperature $T_\mathrm{EW}\sim 100$ GeV. This possibility has been suggested in Weinberg's analysis of gauge symmetries at finite temperature~\cite{Weinberg:1974hy}, and subsequently followed up by several authors\cite{Mohapatra:1979qt,Mohapatra:1979vr,Langacker:1980kd,Hammerschmitt:1994fn,Dvali:1995cj,Dvali:1996zr,Cline:1999wi,Patel:2012pi,Patel:2013zla,Blinov:2015sna}
%{\color{magenta} add Senjanovic, Schmidt references} {\color{blue} check} \cite{Mohapatra:1979qt,Mohapatra:1979vr,Dvali:1995cj,Dvali:1996zr}.  
As observed in Ref.~\cite{Weinberg:1974hy}, for example, Rochelle salt has the remarkable property first undergoing a symmetry-breaking transition as the temperature is lowered, followed by a symmetry-restoring transition at lower temperature\cite{RochelleRef}.
This raises the fascinating possibility that a similar phenomenon may occur in gauge theories\cite{Weinberg:1974hy}.

In light of this possibility and the constraints on the EWBG paradigm, we consider a multi-step phase transition 
beginning with a symmetric universe at high temperature, followed by the spontaneous breaking of SU(3)$_C$ as 
the Universe cools and ending with its subsequent restoration. 
Although there have been studies of multistep phase transitions incorporating 
SU(3)$_C$-breaking~\cite{Laine:1998jb,Cline:1999wi,Patel:2013zla}, only the last has gives a viable mechanism to 
break SU(3)$_C$ symmetry and restore it at zero temperature\footnote{We note that ref \cite{Patel:2013zla} did not analyse the strength of the phase transition or which parts of the parameter space have sufficiently fast tunnelling. A detailed investigation into color breaking phase transitions is the subject of ongoing research.
}. 

In this study, we follow the general set-up of Ref.~\cite{Patel:2013zla} where SU(3)$_C$-breaking is induced by colored scalars obtaining a vacuum expectation value (vev) during the first transition, which breaks both the color SU(3)$_C$ and and electroweak (EW) SU(2)$_L\times$U(1)$_Y$ symmetries of the SM. This vev is then erased during the subsequent transition to the present \lq\lq Higgs phase", wherein the only the neutral component of the Higgs doublet obtains a vev. We refer to these two transitions as the CoB and EW phase transitions, respectively (though technically both break EW symmetry).
We develop a full working scenario of baryogenesis under these conditions, which we refer to as color-breaking baryogenesis (CoBBG).  
We will focus our attention on the CP violation and charge transport dynamics and not the dynamics of the phase 
transition that was previously studied in Ref.~\cite{Patel:2013zla}.
%We propose a new mechanism for the production of the baryon asymmetry of the Universe which exploits a multistep phase transition in which SU(3) color symmetry is first broken and then restored. During the first transition, SU(3) symmetry is spontaneously broken along with SU(2)xU(1) by a scalar field interacting with Standard Model fermions as a leptoquark. The combination of SU(3) breaking and leptoquark interactions spontaneously breaks baryon number, leading to abundant production of baryons at the beginning of the intermediate phase. The same interactions further produce baryons via a mechanism equivalent to electroweak baryogenesis during the same transition, making the total baryon asymmetry the sum total contribution of the two mechanisms. Washout processes also occur in the intermediate phase but the baryon asymmetry can be made long-lived by suppressing leptoquark couplings and needs only to last until the second transition. During the second transition SU(3) color symmetry is restored while SU(2)xU(1) remains broken, quenching all washout processes and freezing in the existing baryon asymmetry as long as this transition is not first order. We illustrate this mechanism with a simple leptoquark model that reproduces the observed baryon asymmetry and discuss how certain aspects of such a scenario may potentially be probed by future electric dipole moment and collider searches.
%Moreover, if the new scalar couples to Standard Model (SM) fermions as a leptoquark, the first transition also spontaneously breaks global B-L charge.

To demonstrate this new paradigm, we introduce two new scalar fields, C$_{1,2}$, that are charged under SU(3)$_C$ as well as SU(2)$_L \times$U(1)$_Y$. In order to prevent the existence of stable colored relics, we take these fields to interact with the Standard Model (SM) as  leptoquarks through Yukawa-type interactions. With this field content, the thermal history of symmetry breaking is
\bea
\nonumber
SU(3)_C\times SU(2)_L\times U(1)_Y\\
\nonumber
\overset{T_1}{\longrightarrow} SU(2)_C\times U(1)_{X_1}\times U(1)_{X_2}\\
\overset{T_2}{\longrightarrow} SU(3)_C\times U(1)_{EM}\ \ \ ,
\eea
where $X_{1,2}$ denote two independently conserved U(1) charges during the CoB phase that accompany a residual color SU(2)$_C$ symmetry. The BAU is generated during the first phase transition at temperature $T_1$, with the Sakharov conditions realized as follows:
\begin{itemize}
\item[1.] Baryon number conservation is violated in two ways: the usual electroweak sphalerons anomalously violating $B+L$ and spontaneous violation of $B-L$ in the color breaking phase, since the leptoquark fields $C_j$ carry $B-L$.
\item[2.] The leptoquark-quark-lepton Yukawa couplings contain  new CP-violating complex phases that source the generation of charge asymmetries during the first transition.  
%We discuss the consequences for experiments searching for permanent electric dipole moments in Section~\ref{sec:Pheno}.
\item[3.] The spontaneous breaking of SU(3)$_C$ symmetry proceeds via a strongly first order phase transition, resulting in nucleation of CoB bubbles and, thereby, satisfying the out-of-equilibrium requirement.
\end{itemize}
During the second transition at temperature $T_2$, the BAU produced during the first step inside the CoB phase is transferred to the Higgs phase. So long as the second transition does not permit re-excitation of the unbroken phase EW sphalerons or injection of significant entropy, the first phase BAU will not be washed out or diluted when the second transition occurs. 

During the first step, the BAU produced via electroweak sphalerons is directly analogous to EWBG. Electroweak sphalerons are unsuppressed in the symmetric phase. CP violating interactions with the walls of the expanding CoB phase bubbles creates a total left handed number density that biases the sphalerons at the bubble exteriors. This produces a net $B+L$ asymmetry, some of which is swept up by the advancing bubble wall. For a sufficiently strong first order CoB transition, the broken phase EW sphaleron transitions will be sufficiently quenched by the $C_j$ vevs so as to preclude washout of the $B+L$ asymmetry.

The second mechanism for violating baryon number conservation involves the spontaneous violation of $B-L$ 
number conservation by the $C_j$ vevs.   %This occurs alongside a spontaneous violation of electric charge 
%($Q_\mathrm{EM}$) conservation. The relic $Q_\mathrm{EM}$ and $B-L$ asymmetry resulting from this mechanism 
%is spacetime dependent and quickly relaxes to a negligible value at a scale that is typically a trillionth of the Hubble length 
%for our model.  During the second transition, these charge asymmetries are also transferred to the present Higgs phase. 
%As we show below, the resulting relic $Q_\mathrm{EM}$ to entropy ratio  is may orders of magnitude below the present 
%constraints derived from the isotropy of the cosmic microwave background\cite{Caprini:2005bh}. 
%The corresponding relic $B-L$ asymmetry 
%is also negligible.
The total $B-L$ inside and outside the bubble is zero, however a non-zero $B-L$ density is trapped inside the expanding bubble. The size of this contribution however is negligible as the $B-L$ density relaxes to zero within a trillionth of the Hubble length at the time of nucleation and will continue to diffuse. On the other hand, 
the $B+L$ asymmetry is effectively conserved deep within the color broken phase and persists into the 
electroweak phase. The net BAU is, thus, dominated by the conventional $B+L$ generating EWBG mechanism. 
We find that given the present phenomenological constraints from collider searches and EDMs, the resulting BAU 
can be comparable in magnitude to the observed asymmetry.

We organize our discussion of this scenario as follows. In the Section \ref{sec:Model} we define our exact choice of model to illustrate this scenario and section~\ref{sec:SSB} elaborates on the symmetry breaking patterns associated with the multi-step phase transition. In Section~\ref{sec:CoBBG} we analyze all issues of charge transport including local equilibrium considerations, derivation of quantum transport equations, and our results including the contribution from the electroweak mechanism. Section~\ref{sec:Pheno} discusses the zero temperature phenomenology before we conclude in Section~\ref{Conclusions}.%Section~\ref{sec:charge} discusses the relic charge asymmetry and observational bounds before discussing the zero temperature phenomenology of such a model of Section~\ref{sec:Pheno}. Finally, we conclude in Section~\ref{Conclusions}.

\section{The Model}\label{sec:Model}
%%%%%%%%%%%%%%%%%%%%%%%%%%%%%%%%%%%%%%%%%%%%%%%%%%%%%%%%%%%%%%%%%%

Our illustrative model consists of the SM plus two scalar leptoquark fields, $C_{1,2}$, that must be charged under SU(3)$_C$ and SU(2)$_L$ in order to catalyze a CoB phase transition and quench electroweak sphalerons during this transition. In general, there are three scalar leptoquark representations that couple to SM fermions and have non-trivial SU(2)$_L$ quantum numbers~\cite{Agashe:2014kda}, (3,3)$_{-1/3}$, (3,2)$_{7/6}$, and (3,2)$_{1/6}$. We seek a model that has the minimum number of free parameters, is least constrained phenomenologically, and does not enable any baryon number violating processes at zero temperature. 

The (3,3)$_{-1/3}$ representation does not pass these requirements as it couples to both $Q_LQ_L$ and $Q_L^\dagger L_L^\dagger$, violating baryon number explicitly at tree level. %Even restricting oneself to third generation bilinears, the $p \to K^+ \bar{\nu}$ decay mode appears at the 1-loop level, imposing severe constraints~\cite{Baldes:2011mh}. 
The (3,2)$_{7/6}$ representation admits an enhanced $\mu \to e \gamma$ rate by virtue of it coupling to both $Q_L^\dagger e_R$ and $u_R^\dagger L_L$. The same enhancement also appears in 1-loop logarithmically divergent contributions to the charged lepton mass matrix, leading to non-trivial naturalness constraints even if the leptoquarks only couples to third generation particles~\cite{Arnold:2013cva}. In contrast, the (3,2)$_{1/6}$ representation only couples to $d_R^\dagger L_L$, so it is not subject to the above phenomenological constraints. Furthermore it has no perturbative baryon number violation and can catalyze gauge coupling unification \cite{Murayama:1991ah}. We therefore use this representation to illustrate the mechanism of CoBBG. \\

Given the quantum numbers of the $C_{1,2}$ fields, we denote them as 
\bea
C_{j \alpha} = \left( 
\begin{array}{c}
\chi_{j \alpha}^{2/3}  \\
\chi_{j \alpha}^{-1/3}
\end{array}
\right)
\qquad j = 1,2
\eea
where Roman and Greek subscripts indicate the field and SU(3)$_C$ indices respectively, while superscripts represent the electromagnetic charges of the SU(2)$_L$ component fields. The most general Yukawa interaction for this representation is
\bea
\mathscr{L}_{Y} \supset \overline{d}_{R}^\alpha \left(  Y^1 C_{1 \alpha}+Y^2 C_{2 \alpha}  \right) L  + h.c.
\label{eq:yuk_ints}
\eea
where flavor indices have been suppressed and the SU(2)$_L$ contraction is \begin{equation} C_{j \alpha} L \equiv \chi_{j \alpha}^{2/3} e_L - \chi_{j \alpha}^{-1/3} \nu_L\ .\end{equation} The 3$\times$3 Yukawa matrices, $Y^1$ and $Y^2$, couple right handed down-type quarks ($d_R, s_R, b_R$) to left handed leptons ($e_L, \mu_L, \tau_L$) and neutrinos ($\nu_{e L}, \nu_{\mu L}, \nu_{\tau L}$). These new Yukawa matrices are arbitrary. However, the absence of large flavor changing neutral currents (FCNC) combined with LHC constraints being significantly more stringent for leptoquarks coupled to first or second generation fermions~\cite{Aad:2015caa} suggests a hierarchical structure where $\overline{b}_R$-$\nu_{\tau L}$ couplings are dominant. Consequently, we take
\bea
Y^i = \text{diag} \left( 0, 0, \tilde{y}_i \right) ,
\label{eq:yuk_texture}
\eea
where the zeros here indicate sub-leading couplings that we neglect in our analysis. Consequently, there is only one rephasing invariant CP phase and it is the relative phase Im($\tilde{y}_1^* \tilde{y}_2$). This is the minimum structure necessary to illustrate the CoBBG mechanism.  

Note that in Ref.~\cite{Patel:2013zla} it was found that in order to have a phenomenologically viable scenario 
where color symmetry is broken and restored, one requires there to be gauge singlets in the model. 
The gauge singlet allows the leptoquark mass to be a TeV or higher and can result in the leptoquark mass during the
color breaking phase to substantially differ from its zero temperature value in the standard model phase. In this
paper we ignore gauge singlets and leave such features to future work.
%where the dynamics of the charged transport
%and the phase transition are simultaneously studied. {\color{blue} This appears to be the right place to put this. Do you agree?} {\color{magenta} the foregoing discussion belongs elsewhere, where we discuss the EWPT}

%%%%%%%%%%%%%%%%%%%%%%%%%%%%%%%%%%%%%%%%%%%%%%%%%%%%%%%%%%%%%%%%%%
\section{Spontaneous Symmetry Breaking of SU(3)$_C \times$SU(2)$_L \times$U(1)$_Y$}\label{sec:SSB}
%%%%%%%%%%%%%%%%%%%%%%%%%%%%%%%%%%%%%%%%%%%%%%%%%%%%%%%%%%%%%%%%%%

In this section, we describe the pattern of spontaneous symmetry breaking during the multistep phase transition. It is necessary to understand this pattern as the presence of conserved gauge symmetries during the CoB phase will be used to make significant simplifcations in the next section. 

Without loss of generality, we choose the orientation of the color breaking vevs such that the shifted $C_j$ fields are
\bea
\left( 
\begin{array}{c}
\chi_{j \alpha}^{2/3}  \\
\chi_{j \alpha}^{-1/3}
\end{array}
\right)
\to
\left( 
\begin{array}{c}
\chi_{j \alpha}^{2/3}  \\
\varphi_j \delta^3_\alpha + \sigma_{j \alpha}^{-1/3}
\end{array}
\right)
\qquad j = 1,2
\eea
where $\varphi_j$ are the vevs and the $\delta^3_\alpha$ singles out a direction in SU(3)$_C$ space. In order to identify the symmetry breaking pattern, we examine the gauge boson mass spectrum in the CoB phase.

Neglecting fluctuations around the vevs, the gauge boson mass spectrum is given by 
\begin{eqnarray}
&& \sum_j \left| D_\mu \langle C_{j \alpha} \rangle \right|^2  = \varphi_{CB}^2  \nonumber \\
&&  \times \bigg[ e_S^2 \left( G^{+,45}_\mu G^{-,45 \mu} + G^{+,67}_\mu G^{-,67 \mu} \right) 
 + e_W^2 W^+_\mu W^{- \mu}  \nonumber \\ &&+  \ds  \left( 
\begin{array}{ccc}
W^3_\mu & B_\mu & G^8_\mu 
\end{array}
\right)  \frac{\mathcal{M}^2}{2} \left(
\begin{array}{c}
W^{3 \mu} \\
B^\mu \\
G^{8 \mu}
\end{array}
\right)  \bigg] \nonumber \\
\label{eq:masses}
\end{eqnarray}
where $\varphi_{CB}^2 \equiv \varphi_1^2 + \varphi_2^2$ and the hypercharge, weak, and strong gauge couplings have been normalized as \begin{equation}
(e_Y, e_W, e_S) \equiv (g_Y/\sqrt{2}, g_W/\sqrt{2}, g_S/\sqrt{2}) \ .
\end{equation} The $G^{\pm, ij}_\mu \equiv \frac{1}{\sqrt{2}} \left( G^i_\mu \mp i G^j_\mu \right)$ fields correspond to the well-known SU(3)$_C$ generators of isospin ($ij$=$12$), U-spin ($ij$=$45$), and V-spin ($ij$=$67$) while the $W^\pm \equiv \frac{1}{\sqrt{2}} \left( W^1_\mu \mp i W^2_\mu \right)$ fields correspond to the familiar generators of weak isospin. 

The 3$\times$3 mass matrix takes the form
\bea
\hspace*{-.2in}\mathcal{M}^2 = 
\left( 
\begin{array}{ccc}
\ds e_W^2 & \ds - \frac{e_W e_Y}{3} & \ds \frac{2 e_W e_S}{\sqrt{3}} \vspace*{.05in} \\
\ds -\frac{e_W e_Y}{3} & \ds \frac{e_Y^2}{9} & \ds - \frac{2}{3 \sqrt{3}} e_S e_Y     \vspace*{.05in} \\
\ds \frac{2 e_W e_S}{\sqrt{3}} & \ds  - \frac{2}{3 \sqrt{3}} e_S e_Y   & \ds \frac{4}{3} e_S^2
\end{array}
\right) .
\eea
This matrix has only one non-zero eigenvalue, implying the presence of two unbroken and one broken U(1) gauge symmetries present in the CoB phase. We denote the corresponding three mass eigenstate fields as 
\be
\left(X_{1\mu},\ X_{2\mu},\ X_{3\mu}\right)^T = \mathcal{U}\ \left(W^3_\mu,\ B_\mu,\ G^8_\mu\right)^T
\ee
where $\mathcal{U}$ diagonalizes $\mathcal{M}^2$.
%$X_{1\mu}$, $X_{2\mu}$, and $X_{3\mu}$, which are mixtures of $W^3_\mu$, $B_\mu$, and $G^8_\mu$ determined by diagonalizing $\mathcal{M}^2$. 
While $X_{1 \mu}$ and $X_{2 \mu}$ remain massless, mediating long-range forces associated with the unbroken symmetries U(1)$_{X_1}$ and U(1)$_{X_2}$, $X_{3 \mu}$ develops a mass 
\be
m_{X_3}^2 = \varphi_{CB}^2 /9 \left( 12 e_S^2 + 9 e_W^2 + e_Y^2 \right)
\ee
and thus mediates a short-range force associated with the broken U(1)$_{X_3}$ symmetry. The corresponding charge generators of these U(1) symmetries are given by
\begin{align}
&Q_{X_1} = T^8 - \ds \frac{2}{\sqrt{3}} \tau^3 \nonumber \\
&Q_{X_2} = \tau^3 + 3 Y \nonumber \\
&Q_{X_3} = T^8 + \ds \frac{\sqrt{3}}{2} \frac{e_W^2}{e_S^2} \tau^3 - \frac{1}{2 \sqrt{3}} \frac{e_Y^2}{e_S^2} Y \ .
\end{align}

\begin{figure}[t!]
\centering
\includegraphics[scale=.5]{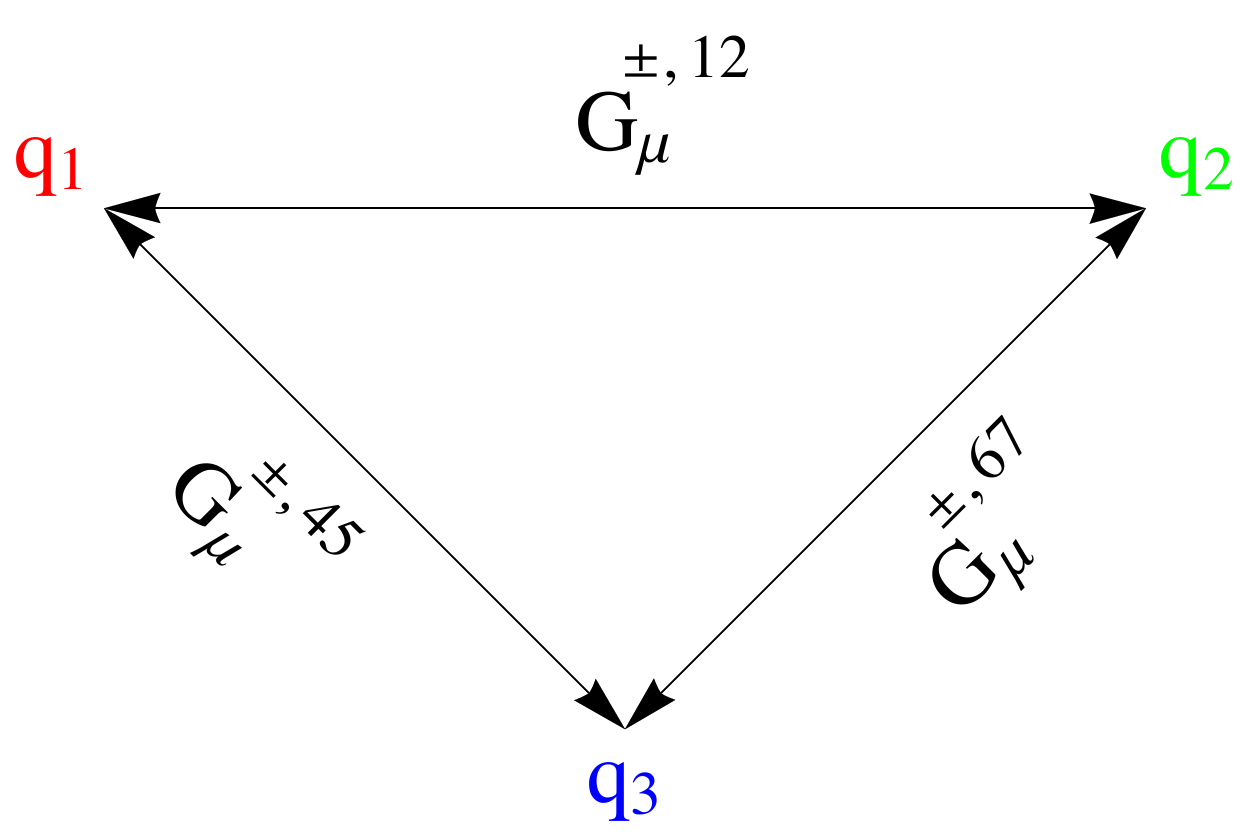}
\caption{The interaction pattern of gauge fields corresponding to SU(3)$_C$ generators of isospin ($G^{\pm, 12}_\mu$), U-spin ($G^{\pm, 45}_\mu$), and V-spin ($G^{\pm, 67}_\mu$) and fields in the fundamental triplet representation.}
\label{fig:gluontriangle}
\end{figure}

The charges $Q_{X_1}$ and $Q_{X_2}$ are conserved in both the symmetric and color broken phase. The gauge fields corresponding to the SU(3)$_C$ isospin generators are missing from Eq.~(\ref{eq:masses}) and thus remain massless, indicating the existence of an unbroken SU(2)$_C$ subgroup of SU(3)$_C$ in the CoB phase.  This situation effectively distinguishes color state $q_3$ from $q_1,q_2$, indicating that its' dynamics should be treated separately in the CoB phase. This situation is represented graphically in Fig.~\ref{fig:gluontriangle} and further clarified in Section~\ref{sec:CoBBG1}. In Section~\ref{sec:CoBBG2}, we will study the charge transport dynamics of each independent color separately.
\section{Baryogenesis}\label{sec:CoBBG}
%%%%%%%%%%%%%%%%%%%%%%%%%%%%%%%%%%%%%%%%%%%%%%%%%%%%%%%%%%%%%%%%%%

%In this section, we describe the calculation of the BAU in the CoBBG scenario. 

The BAU calculation is performed in two steps. First, we analyze charge transport dynamics of the relevant number densities in order to calculate the space time varying $B-L$ and chiral charge densities generated during the strongly first-order CoB phase transition. Second, the total left handed number density that biases the sphalerons resulting in a $B+L$ asymmetry via the EWBG mechanism. Combining the results of contributions yields the net BAU in the CoB phase.

The dynamics of particle number densities during a first order phase transition is a highly non-Markovian process that depends on the entire history of the system. In particular, ``memory effects'' can lead to a resonant boost of both CP violating sources and CP conserving relaxation terms  that result from interactions with the space time varying vacuum \cite{Riotto:1998zb,Carena:2002ss,Konstandin:2005cd,Lee:2004we}. Recall that our model contains a new, $T=0$ rephasing invariant  the leptoquark interaction that results in a new CP-violating (CPV) and $(B-L)$-violating (BLV) source $S_i^{(\mathrm{CPV},\, \mathrm{BLV})}$ for the transport equations 
\bea
\partial_\mu j_i^\mu = -  \sum_j \Gamma_{ij} \mu_j + S_i^{(\mathrm{CPV},\, \mathrm{BLV})}
\label{eq:general_boltzmann_form}
\eea
where $j_i^\mu$ and $\mu_i$ are the charge current density and chemical potential, respectively, of particle species $i$ and $\Gamma_{ij}$ are the rates of interactions between species $i$ and $j$. 

\begin{figure}[t]
%\centering
\includegraphics[scale=.5]{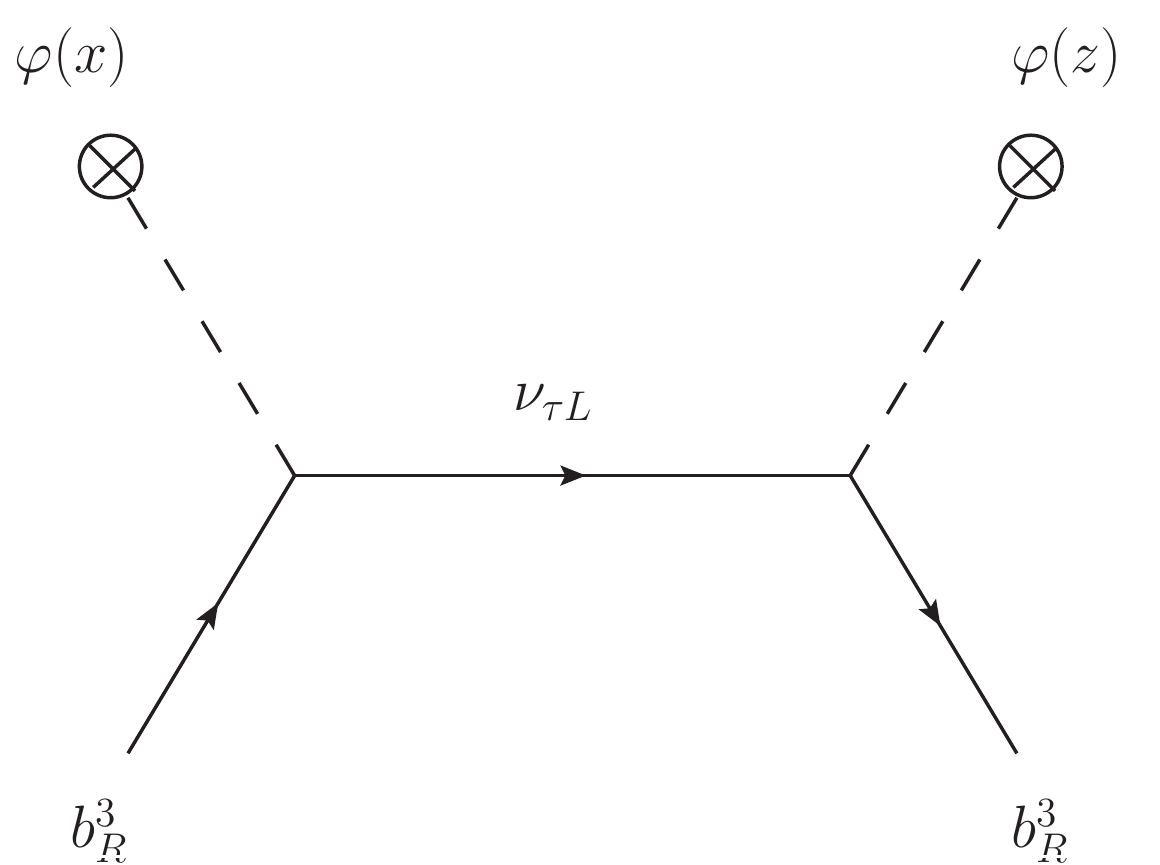}
\caption{Interaction between the left handed tau neutrino and the right handed 3rd color bottom quark with space-time varying vacuum. This interaction is responsible for new CP violating source}
\label{fig:vevinsertion}
\end{figure}

The computation of the $S_i^{(\mathrm{CPV},\, \mathrm{BLV})}$ is, in general, quite subtle, and there remain a number of open theoretical issues for the CPV-sources involving fermions (for a discussion, see {\em e.g.}, Ref.~\cite{Morrissey:2012db} and references therein). The general framework we adopt is the Schwinger-Keldysh closed time path formalism~\cite{Schwinger:1960qe,PhysRev.126.329,Bakshi:1963-1,Bakshi:1963-2,Keldysh:1964ud,CHOU19851}. We will work with the vev insertion approximation (VIA), wherein we treat space-time varying vevs appearing in the $b$-$\nu$ mass matrix perturbatively to lowest non-trivial order. The diagrammatic representation of $S_i^{(\mathrm{CPV},\, \mathrm{BLV})}$ in the VIA is shown Fig.~\ref{fig:vevinsertion}. We expect that the VIA gives a reasonable guide to the magnitude of the CPV effects and allows one to see structure of the dynamics in our scenario. A more refined treatment including full accounting for flavor oscillations and vev-resummations is in progress\cite{Carena:2002ss,Konstandin:2005cd,Cirigliano:2009yt,Cirigliano:2011di}, and it remains unclear as to whether the VIA yields an overestimate or underestimate. Consequently, we will take our results as indicative of the magnitude of the BAU in our set up and not as numerically definitive.

With these caveats in mind, we apply the techniques in Ref.~\cite{Lee:2004we,Liu:2011jh}, we obtain 
\begin{eqnarray}
&& S_b^{(\mathrm{CPV},\, \mathrm{BLV})}= -S_{\nu_L}^{(\mathrm{CPV},\, \mathrm{BLV})} =\nonumber\\  
&&\ds  \frac{{\rm Im} [\tilde{y}_1\tilde{y}_2] }{\pi ^2}v _{CB}^2(z) \frac{\partial \zeta(z)}{\partial t} \int _0 ^\infty \frac{k^2 dk}{ \omega _{\nu _L} \omega _b}   \nonumber \\
&&   {\rm Im} \Bigg[     \left( \mathcal{E}_{\nu _L} \mathcal{E}_b + k^2 \right) \left( \frac{n_f(\mathcal{E}_{\nu _L}) + n_F(\mathcal{E}_b)}{(\mathcal{E}_{\nu _L} +\mathcal{E}_b)^2} \right)  \nonumber \\ 
&&+ \left( \mathcal{E}_{\nu _L} \mathcal{E}_b ^* - k^2 \right) \left( \frac{n_f(\mathcal{E}_{\nu _L}) - n_F(\mathcal{E}_b ^ *)}{(\mathcal{E}_b^*-\mathcal{E}_{\nu _L})^2} \right) \Bigg]   . 
\label{sourceterm}
\end{eqnarray}
Here $\tan \zeta (z)$ is the ratio of the vevs of the colored scalars, $\varphi_2(z)/\varphi_1(z)$, $n_F$ is the  Fermi-Dirac distribution function,  $\omega_i = \sqrt{k^2 + m_i^2}$, and $\mathcal{E}_i \equiv \sqrt{k^2 + m_i^2} - i \Gamma_i$ with $m_i$ and $\Gamma_i$ representing the fully corrected thermal mass\footnote{for a more detailed treatment of thermal masses in phase transitions see \cite{Curtin:2016urg}} and width of state $i$ \footnote{In principle, one can have CP violating sources resulting from CP violation in the scalar potential, e.g., see~\cite{Inoue:2015pza}. However, for the purposes of this paper, we only consider the CP violating source listed in Eq.~(\ref{sourceterm}).}.

Denoting the chemical potentials of the left handed tau neutrino and the third color-component of the right handed bottom quark as $\mu _{\nu _{\tau L}}$ and $\mu _{b_3}$, respectively, we can write the CP conserving relaxation term associated with Fig~\ref{fig:vevinsertion} as 
\bea
S^{CP} = (\mu _{\nu _{\tau L}}-\mu _{b ^3_R} ) \Gamma _M
\label{eq:relaxationterm}
\eea
with
\bea
& \Gamma _M = \ds \frac{|\tilde{y}_1\varphi_1(z)+\tilde{y}_2 \varphi_2(z)|^2}{2 \pi ^2 T}   \int _0 ^\infty \frac{k^2 dk}{\omega _\nu \omega _b} & \nonumber \\ 
&\ds  {\rm Im} \Bigg[ \left( \mathcal{E}_{\nu _L} \mathcal{E}_b + k^2 \right)   \left( \frac{h_F(\mathcal{E}_{\nu _L} )+ h_F(\mathcal{E}_b)}{\mathcal{E}_{\nu _L} +\mathcal{E}_b} \right) & \nonumber  \\ 
& \ds -\left( \mathcal{E}_{\nu _L} \mathcal{E}_b^* - k^2 \right) \left( \frac{h_F(\mathcal{E}_{\nu _L} )+ h_F(\mathcal{E}_b^*)}{\mathcal{E}_b ^*-\mathcal{E}_{\nu _L}} \right)\Bigg] &.
\label{eq:relaxationterm1}
\eea
and
\begin{equation} h_F(x) = e^{x/T} / (1 + e^{x/T})^2. \end{equation} 
\\
%The divergence of each particle number current is related to the list of self energy interactions using the same techniques. The result is a network of transport equations of the form
%

%
With these sources in hand, we now analyze the transport equations (\ref{eq:general_boltzmann_form}) in detail.
A particle's dynamics are important if it is able to diffuse ahead of the advancing bubble wall. 
The diffusion time is characterized by a diffusion constant  $D_i$ (see below) and the bubble wall velocity $v_w$: $\tau_\mathrm{diff} =
D/v_w^2\sim 10^4/T$ \cite{Cohen:1994ss} for $v_w$ on the order of 0.05. This time scale is typically shorter than the inverse rate for the EW sphalerons to convert the left-handed number density $n_L$ into $B+L$, $\tau_\mathrm{EW} \sim \Gamma_\mathrm{EW}^{-1} \sim10^5/T$, where $\Gamma_\mathrm{EW} \approx 120 \alpha_W^5 T$ and $\alpha_W$ is the SU(2)$_L$ fine structure constant\cite{Bodeker:1999gx}. Consequently, we may decouple the equations for $n_L$ and $B+L$ generation to a reasonable approximation. 
%We assume then that third generation quarks and left handed leptons, the Higgs doublet and the colored scalars all have time to diffuse in front of the bubble wall due to fast Yukawa interactions with either the Higgs or the leptoquarks. We will also find that strong sphalerons allow for the right handed charm to diffuse fast enough to contribute to the total left handed number density.
\\

Following Refs.~\cite{Lee:2004we}, we assume a planar bubble wall profile so that charge densities are functions only of their displacement $z=|{\vec x}-{\vec v}_wt |$ from the bubble wall in its rest frame, where ${\vec x}$ is the co-ordinate in the plasma rest frame and where  the $z<0$ ($z>0$) region corresponds to the (un)broken phase. We also apply Fick's law to make the replacement $j_i^\mu \equiv (n_i, \vec{j_i}) \to (n_i, - D_i \vec{\nabla} n_i)$. Here, $n_i$ is the charge density and $D_i$ is the diffusion constant which describes how $n_i$ is transported away from the bubble wall. Assuming $\mu_i/T << 1$, the chemical potentials are related to the charge densities as $n_i = (T^2/6) k_i \mu_i + \mathcal{O}(\mu_i/T)^3$ where the $k_i$ factor counts the effective degrees of freedom of species $i$ in the plasma. These $k_i$ factors are
\bea
k_i = g_i \frac{6}{\pi^2} \int_{m_i/T}^\infty dx x \frac{e^x}{(e^x \pm 1)^2} \sqrt{x^2 - m_i^2/T^2}
\eea
where $g_i$ counts the number of internal degrees of freedom for species $i$ and $m_i$ is the effective mass of particle $i$ at temperature $T$. By searching for steady-state solutions that only depend on $z$, we can make the replacements $\partial n_i(z) / \partial t \to v_w n_i'(z)$ and $\vec{\nabla}^2 n_i(z) \to n_i''(z)$ where the prime denotes differentiation with respect to $z$ and $v_w \equiv \partial z / \partial t$ is the velocity of the bubble wall. After these modifications, the Boltzmann equations become a coupled set of second order differential equations for the charge densities $n_i(z)$ with one such equation for each independent particle species coupled to the baryon- and CP-violating source in the plasma. \\
%%%%%%%%%%%%%%%%%%%%%%%%%%%%%%%%%%%%%%%%%%%%%%%%%%%%%%%%%%%%%%%%%%
\subsection{Local Equilibrium Considrations}\label{sec:CoBBG1}
%%%%%%%%%%%%%%%%%%%%%%%%%%%%%%%%%%%%%%%%%%%%%%%%%%%%%%%%%%%%%%%%%%

%In general, the presence of broken SU(3)$_C$ and EW symmetries in the CoB phase increases the number of transport equations that must be considered relative to the symmetric phase 
%{\color{magenta} is this really true? The fluctuations around the colored vevs carry no charge so we are just left with the degrees of freedom charged under SU(2)$_C$ and the two U(1)$_X$ symmetries}. 
%breaking color symmetry naively increases the number of transport equations we have to solve by a factor of three which would make calculations a formidable task. 
%Furthermore we cannot simply decouple the dynamics of color and isospin singlets from multiplets by fiat. In Eq. (\ref{eq:relaxationterm}) for example we have chemical potentials that are components of a SU(2)$_L$, $\mu _{\nu _{\tau L}}$, and a color triplet $\mu _{b_3}$ 
%{\color{magenta} I don't understand what this last sentence says}. {\color{blue}  
%took a sledge hammer to the above paragraph and simply added a couple of short sentences to the next}

The spontaneous breaking of SU(3)$_C$ symmetry implies that one needs to consider the transport dynamics of each
color separately. A significant simplification can be made if the dynamics of colou and weak isospin singlet can
be separated from multiplets. This is what we endeavor to achieve in this section. The only assumption we will require is that
we are in a section of parameter space where gauge interactions are fast enough compared to the inverse of the 
diffusion rate which is controlled by the velocity of the advancing bubble wall, $\Gamma_D\sim v_W^2/D$\cite{Chung:2008aya}.
% As we show below, if we assume that gauge interactions are fast compared to the diffusion rate 
%{\color{magenta} why is this a valid assumption?}{\color{blue}I said to say something like the above and you agreed
%please confirm/deny},
%the residual symmetries of the CoB phase allow us to consider only the transport equations for color and weak isospin singlets. 
%we can make use of the conservation laws that hold in the color broken phase to equate linear combinations of chemical potentials in such a way where color and isospin multiplets can be written in terms of singlets. This allows us to just consider the network of transport equations for color and isospin singlets - a considerable simplification. 
%Note that since the conservation laws are exact, the only approximation is that gauge interactions result in local equilibrium between massive gauge bosons and gauge multiplets.  
For the sake simplicity we will also assume that scalar interactions involving both colored scalars are also fast enough compared
to a diffusion rate to equilibrate the two 
%{\color{magenta} why is this assumption valid?} {\color{blue} I made a moderate
%change here to avoid repetition from the previous sentence}. As such the set of color and isospin singlets we will be left with are
%
\begin{align}
n_Q   &= \sum _i n_{t_{Li}} +n_{b_{Li}}\nonumber \\
n_{T} &= \sum _i n_{t_{Ri}} \nonumber \\
n_{B} &= \sum _i n_{b_{Ri}} \nonumber \\
n_{U} &=\sum _i n_{c_{Ri}} \nonumber \\
n_L &= n_{\tau _L } +n_{\nu _{\tau L}} \nonumber \\
n_H &= n_{H^+} + n_{H^0}\nonumber  \\
n_{C} & = \frac{1}{2} \left( \sum _{i\alpha} n_{\chi _{i\alpha}^{2/3}}+n_{\chi _{i\alpha}^{-1/3}} \right)  \label{eq:independent_set}
\end{align}
where the above species are the left handed third generation quark doublet, the right handed top, bottom and charm, the third generation left handed lepton doublet, the Higgs doublet and the combined colored scalar densities respectively.
Note that $i\in(1,2,3)$ is an SU(3)$_C$ index and $\alpha \in (1,2)$ is an index 
for the species of leptoquark. 
%{\color{blue} This seemed to be the best place for this suggestion. Do you agree?}

Let us begin with making use of gauge interactions. We will denote the chemical potentials of the two components of an arbitrary SU(2)$_L$ doublet as $\mu _{\uparrow}$ and $\mu _{\downarrow}$ respectively. Also let us denote the three components of an arbitrary color triplet as $\mu _i$ for $i \in 1,2,3$.
The remaining SU(2)$_C$ symmetry results in a local equilibrium relation between the first two colors
\begin{equation}
    \mu _1 = \mu _2 \ . 
\end{equation}
The result of this is that there are only two independent colors. We can therefore write all components of the SU(3)$_C$ triplet can be written as a linear combination of the color singlet and octet state which we denote as $\mu _8$ and $\mu  _S$ respectively 
%{\color{magenta} I don't understand the meaning of this equation. What is the $\mu_{1,2}$ on the RHS? The some of both? their average? } 
%{\color{blue} explained below the following equation}
%
\bea 
\left( 
\begin{array}{c}
\mu_S \\
\mu_{8}
\end{array}
\right)
=
\left(
\begin{array}{cc}
2 & 1 \\
\frac{1}{\sqrt{3}} & -\frac{1}{\sqrt{3}}
\end{array}
\right)
\left(
\begin{array}{c}
\mu_{1,2} \\
\mu_{3}
\end{array}
\right) \label{eq:octetbasis}
\eea
where $\mu _{1,2}$ represents either color since they are in equilibrium. 
\\
The assumption of local gauge equilibrium for massive gauge results in the following relations between chemical potentials
%\begin{equation}
\begin{align}
\mu _1 - \mu _3 &=\frac{1}{\sqrt{3}}\mu _8= \mu _{G_{45}} \nonumber \\
\mu _2 - \mu _3 &=\frac{1}{\sqrt{3}}\mu _8= \mu _{G_{67}}  \nonumber \\
\mu _{\uparrow} - \mu _{\downarrow} & \equiv \Delta \mu= - \mu _W \ \ \ ,
\label{eq:chemeq}
\end{align}
where $\mu _{\uparrow}$ ($\mu _{\downarrow}$) denotes the chemical potential for any weak isodoublet with third component $+1/2$ ($-1/2$) .
%\end{equation}
The first two lines in Eq.~(\ref{eq:chemeq})  imply 
\begin{equation}
   \mu _{G_{45}} = \mu _{G_{67}}\equiv \mu_G \label{eq:massiveG}
\end{equation}
%equate the chemical potentials of all color octets to the chemical potential  of a single massive gluon, which can be either $ \mu _{G_{45}}$ or $ \mu _{G_{67}}$ and we denote simply as $\mu_G$ 
%{\color{magenta} do you mean $ \mu _{G_{45}} = \mu _{G_{67}}\equiv \mu_G$ ? If so, explain why}{\color{blue} suggestion
%given at conference implemented above} . 
Similarly any chemical potential of the form $\Delta \mu$ is equal to $- \mu _W$. 
%of all SU(2)$_L$ triplets 
%{\color{magenta} this is quite imprecise. Do you mean the lepton and quark triplets? And is it true only for the $\tau_3$ component of the triplet?}  
%{\color{blue} I think the most efficient thing to do is to give these chemical potentials a name. I imitated your
%notation from the supergauge paper. Is this better?}
%are equal to the chemical potential of the $W^-$ boson. 
Therefore, the densities of all gauge multiplets in the network of transport equations for color and SU(2)$_L$ singlets can be written in terms of massive gauge bosons densities.

Next, we use the CoB phase conservation laws to eliminate the massive gauge boson densities {} from all transport equations for color and SU(2)$_L$ singlets. Recall from section \ref{sec:SSB} that there are the two charges $Q_{X_{1,2}}$ are conserved in the CoB phase. 
%The first, $Q_{X_1}$, is a linear combination of the total color octet charge and the SU(2)$_L$ triplet charge. 
To make use of a conservation law one must set the sum of the charge asymmetry for all particle species to zero. For example, in the case of the $Q_{X_1}$ conservation we have
\begin{equation}
\sum _{i \in {\rm particles}} Q_{X_1}\left(  \frac{6 n _i}{T^2} \right) =\sum _{i \in {\rm particles}} Q_{X_1} \mu _i k_i = 0 \ . 
\end{equation}
For convenience the relevant charges of all important particle species are given in the appendix.  We find for $Q_{X_1}$ the simple relationship
\begin{equation}
\mu _G = -  \mu _{W} \label{eq:wgeq} \ .
\end{equation}
This allows us to eliminate the $\mu _G$ in terms of $\mu _W$.
%chemical potentials of the remaining massive gluon {\color{magenta} please check the grammar and agreement of this sentence. Also do yo mean eliminate $\mu_G$ in terms of $\mu_W$?}. 
%{\color{magenta} please check the grammar and agreement of this sentence. Also do yo mean eliminate $\mu_G$ in terms of $\mu_W$?}. 
%{\color{blue} I think this shorter sentence improves it }
Next we consider $Q_{X2}$ conservation. Using Eqs. (\ref{eq:chemeq}), (\ref{eq:massiveG}) and (\ref{eq:wgeq}) we obtain
% contains the operator $\tau ^3$ it can be used to eliminate the chemical potential for $\mu _W$, thereby achieving our goal of writing all color and isospin multiplets in terms of singlets.  We also assume the total baryon number is locally conserved for the first two generations of quarks.  With these assumptions, we find that setting the total $Q_{X2}$ to zero gives
\begin{eqnarray}
&&  \mu_W = \nonumber  \\ && \frac{3}{16} \left( \frac{1}{6} \mu_{Q_{ L}}  + \frac{2}{3} \mu_{t_R} - \frac{1}{3} \mu_{b_R} - \frac{1}{2} \mu_{L} + \mu_H + \frac{2}{3} \mu_C \right),  \nonumber \\
\label{eq:x2_conservation_result}
\end{eqnarray}
which can be used to eliminate $\mu _W$. We have now achieved our goal of 
writing $\mu _8$ and $\Delta \mu $ in terms of gauge singlet densities.
%{\color{magenta} the logic here is {\bf really opaque}. How does (\ref{eq:x2_conservation_result}) allow one to decouple the singlet and non-singlet equations? }
%{\color{blue} I tried to folow your advice is this clearer? }

There exists one additional relationship that allows us to eliminate one more chemical potential. In the CoB phase, the scalar fluctuations about the CoB VEVs are real scalars that can no longer carry any charge, implying vanishing of their chemical potentials, 
\begin{equation} \mu_{\chi_{\alpha 3}^{-1/3}} = 0\end{equation} with $\alpha \in (1,2)$ denoting the leptoquark species. 
 Using equations (\ref{eq:octetbasis}), (\ref{eq:chemeq}), (\ref{eq:massiveG}) and (\ref{eq:wgeq}) we can derive the 
 relation
\bea
\mu_C = - 7 \mu_W \ \ \ .
\label{eq:WC_relation}
\eea
Substituting into Eq.~(\ref{eq:x2_conservation_result}) and solving for $\mu_C$ allows us eliminate the leptoquark chemical potential in favor of the quark, lepton, and Higgs chemical potentials appearing in Eq.~~(\ref{eq:x2_conservation_result}). 
Thus, the final set of Boltzmann equations need not contain either $\mu_W$ or $\mu_C$.
%an equation for the leptoquark density. Specifically this is achieved through substituting the above 
%equation into the right hand side of Eq. (\ref{eq:x2_conservation_result}) and rearranging to isolate $\mu _W$. {\color{blue} Is this better? I could explicitly write the
%equation but you get some ugly fractions} %{\color{magenta} chemical potential?} {\color{blue} fixed }. \\

To conclude this section we briefly comment on the strong sphaleron rate. Strong sphalerons transitions convert left handed
quarks into right handed quarks and vice versa. Since we are breaking SU(3)$_C$ through a strongly first order phase transition,
the sphaleron rate for the third color 
gets supressed in the CoB phase by a factor controlled by the sphaleron energy, $\Gamma _{\rm sph} \sim \exp[-E_{\rm Sph}/T] $, where
the sphaleron energy itself is proportional to the color breaking vev, $v_{\rm cb}$. Therefore we can ignore strong sphaleron transitions for the 
third color. 
%{\color{magenta} Again, {\bf very opaque}. Need to state that strong sphalerons convert L into R quarks and vice versa; that in the CoB phase, this only happens for the residual SU(2)$_C$ sphalerons; explain why this is true (first order CoB transition) in direct analogy with quenching of the EW sphalerons. \lq\lq vev suppressed" is lingo that is not defined and jargon that a reader may not understand precisely.} 
%{\color{blue} is this better? }
The linear combination of chemical potentials that multiply the remaining SU(2)$_C$ sphaleron rate $\Gamma^{(2)}_{SS}$ in the transport equations is just
\begin{equation}
\mu _{L1}+\mu _{L2}-\mu_{R1}-\mu_{R2} \ .
\label{eq:su2sph}
\end{equation}
Using Eqns. (\ref{eq:octetbasis}) and (\ref{eq:chemeq}), one finds that the contributions from the color octets cancel. Recall that we assume that local baryon number is conserved for the first two generations of particles. The result is that the Eq.~(\ref{eq:su2sph}) can be written in the form 
\begin{equation}
\mu _{L i}-\mu_{Ri}=2 \left( 8 \mu _U + \mu _T + \mu _B - 2 
\mu _Q \right) \ .
\end{equation}

%%%%%%%%%%%%%%%%%%%%%%%%%%%%%%%%%%%%%%%%%%%%%%%%%%%%%%%%%%%%%%%%%%
\subsection{Quantum Transport Equations}\label{sec:CoBBG2}
%%%%%%%%%%%%%%%%%%%%%%%%%%%%%%%%%%%%%%%%%%%%%%%%%%%%%%%%%%%%%%%%%%

We now derive the Boltzmann equations for all relevant color and SU(2)$_L$ singlets. 
To that end, we first construct the Boltzmann equations for the color and isospin components of each field, 
adding them together to obtain the equations for the color and SU(2)$_L$ singlet densities. 
To illustrate, consider the right handed $b$-quark singlet charge density $n_B$. 
Following the steps laid out in the previous subsection %{\color{blue} missing word added} {\color{magenta} previous what? Which steps? Unclear}, 
we obtain the following equations for the two independent charge densities $n_{b_R^1}$ and $n_{b_R^3}$
\begin{subequations}
\bea
& \partial_\mu j_{b_R^1}^\mu = - \ds   2 \Gamma_{\chi^{2/3}} \left( \mu_{b_R^1} - \mu_{\chi_{1}^{2/3}} - \mu_{\tau_L} \right) & \nonumber \\
\label{eq:nb1}
& -  2 \Gamma_{\chi^{-1/3}} \left( \mu_{b_R^1} - \mu_{\chi_{1}^{-1/3}} - \mu_{\nu_{\tau L}} \right) &  \\
& -  \; \Gamma_{SS}^{(2)} \sum\limits_{\substack{i=gen.}} \left( \mu_{u_{i R}^1} + \mu_{d_{i R}^1} - \mu_{u_{i L}^1} - \mu_{d_{i L}^1}  \right) , & \nonumber \\
& \partial_\mu j_{b_R^3}^\mu = - \ds 2 \Gamma_{\chi^{2/3}} \left( \mu_{b_R^3} - \mu_{\chi_{3}^{2/3}} - \mu_{\tau_L} \right) & \nonumber \\
& -  2 \Gamma_{\chi^{-1/3}} \left( \mu_{b_R^3} - \mu_{\nu_{\tau L}} \right) - \Gamma_{M} \left( \mu_{b_R^3} - \mu_{\nu_{\tau L}} \right)  &  \nonumber \\
\label{eq:nb3}
& + S^{(\mathrm{CPV},\, \mathrm{BLV})} &
\eea
\end{subequations}
where $\Gamma_{\chi^{2/3}}, \ \Gamma_{\chi^{-1/3}}$ are the 3-body rates stemming from the Yukawa interactions in Eq.~(\ref{eq:yuk_ints}) and $\Gamma_{SS}^{(2)}$ is the strong sphaleron rate associated with non-perturbative SU(2)$_C$ gauge interactions. 

Since the gluons associated with SU(2)$_C$ only mediate interactions between the first two components of a SU(3)$_C$ triplet, the strong sphaleron interactions connected with SU(2)$_C$ have no effect on any charge densities corresponding to the third color. Both the 2-body CP-conserving rate $\Gamma_M$ and the baryon- and CP-violating source term $S^{(\mathrm{CPV},\, \mathrm{BLV}}$ originate from the interactions with the CoB vev and thus only appear in Eq.~(\ref{eq:nb3}). Moreover, the chemical potential $\mu_{\chi_{3}^{-1/3}}$ has vanished due to the formation of CoB VEVs. As a consequence, the combination $\mu_{b_R^3} - \mu_{\nu_{\tau L}}$ is relaxed by both 3-body, $\Gamma_{\chi^{-1/3}}$, and 2-body, $\Gamma_M$, interaction rates in Eq.~(\ref{eq:nb3}). The factors of two in front of the $\Gamma_{\chi^{2/3}}$ and $\Gamma_{\chi^{-1/3}}$ rates represent the contributions from both doublets $C_1$ and $C_2$ whose individual isospin components have been equilibrated by potential operators.\\ 

Before taking the singlet combination of Eqs.~(\ref{eq:nb1},\ref{eq:nb3}), we simplify them by assuming that all $\chi$ fields have the same mass, implying that all 3-body rates are equal up to the (relatively negligible) difference between the $\tau_L$ and $\nu_{\tau L}$ thermal masses, \ie $\Gamma_{\chi^{2/3}} = \Gamma_{\chi^{-1/3}} \equiv \Gamma_C$. The singlet combination is 
%then obtained by taking $\partial_\mu j_B^\mu \equiv 2 \partial_\mu j_{b_R^1}^\mu + \partial_\mu j_{b_R^3}^\mu$ which yields
%
\bea
 \partial_\mu j_B^\mu\equiv  2 \partial_\mu j_{b_R^1}^\mu + \partial_\mu j_{b_R^3}^\mu &\nonumber \\
=
- \ds \bigg( \frac{1}{6} \left( \Gamma_C + \Gamma_M \right) \left( 2 \mu_B - \mu_C - 3 \mu_L \right) & \nonumber \\
 +\ds  \frac{2}{3} \Gamma_{SS}^{(2)} \left( 8 \mu_U + \mu_T + \mu_B - \mu_Q \right) \bigg) + S^{(\mathrm{CPV},\, \mathrm{BLV}} &
\label{eq:prelim_B_eq}
\eea
where $\mu_C \equiv \sum_\alpha (\mu_{\chi^{2/3}_\alpha} + \mu_{\chi^{-1/3}_\alpha})$. Here, we have taken advantage of all equilibrium relations derived in section~\ref{sec:CoBBG1} to write the right hand side of Eq.~(\ref{eq:prelim_B_eq}) entirely in terms of weak isospin and color singlets. Note that to this point we have factored out a factor of 3(2) %{\color{magenta} what are these factors of 3, 2?} {\color{blue} clarified at the end of this sentence }
from the k-factors to account for the components of color triplets (isospin doublets). We now reabsorb these factors into the k-factors appearing in our Boltzmann equations. The final form then becomes
\bea
& v_w B' - D_q B'' = - \left( \Gamma_C + \Gamma_M \right) \left( \ds \frac{B}{k_B} - \frac{C}{k_C} - \frac{L}{k_L} \right) & \nonumber \\
& + 2 \Gamma_{SS}^{(2)} \left( \ds \frac{8 U}{k_U} + \frac{T}{k_T} + \frac{B}{k_B} - \frac{2 Q}{k_Q} \right)  + S^{(\mathrm{CPV},\, \mathrm{BLV})} ,&
\eea
where we have expressed the left hand side of the Boltzmann equation in terms of the singlet density as described in section~\ref{sec:CoBBG}. \\

Following the steps laid out above for all other independent charge densities in Eq.~(\ref{eq:independent_set}), we obtain
\begin{subequations}
\bea 
 v_w U' - D_q U'' &=& - 2 \Gamma^{(2)}_{SS}  \; \mathcal{E}_{SS}  \label{eq:QTEs1} \\
 v_w T' - D_q T'' &=& - 2 \Gamma^{(2)}_{SS}  \; \mathcal{E}_{SS} - \Gamma_H \; \mathcal{E}_H  \label{eq:QTEs2} \\
 v_w Q' - D_q Q'' &=& 4 \Gamma^{(2)}_{SS}  \; \mathcal{E}_{SS} + \Gamma_H \; \mathcal{E}_H  \label{eq:QTEs3} \\
 v_w H' - D_L H'' &=& \Gamma_H \; \mathcal{E}_H  \label{eq:QTEs4} \\
 v_w B' - D_q B'' &=& - 2 \Gamma^{(2)}_{SS}  \; \mathcal{E}_{SS} - \left( \Gamma_C + \Gamma_M \right) \; \mathcal{E}_M  \nonumber \\ 
 + S^{(\mathrm{CPV},\, \mathrm{BLV})} & \label{eq:QTEs5} \\
 v_w L' - D_L L''& =& \left( \Gamma_C + \Gamma_M \right) \; \mathcal{E}_M - S^{(\mathrm{CPV},\, \mathrm{BLV})}  
 \nonumber \\
\label{eq:QTEs}
\eea
\end{subequations}

{%\color{blue} We discussed this in Leiden. The B+L is clear conservation in the third generation. In the first two generations you have explicit B conservation assumed to derive the list of chem potentials in front of the strong sphaleron rate. So there are probably some hidden symmetries in the eqns. Having said
%that I can't remember what you wanted here? Does the above suffice?}{\color{magenta} How do I see $B+L$ conservation? It is true of for b+c+e+f but not for the remaining quarks}  with linear combinations of charge densities
%
\bea
& \mathcal{E}_H \equiv \left( \frac{T}{k_H} - \frac{Q}{k_Q} - \frac{H}{k_H} \right), \;   \mathcal{E}_M \equiv \left( \frac{B}{k_B} - \frac{C}{k_C} - \frac{L}{k_L} \right) & \nonumber \\
& \mathcal{E}_{SS} \equiv \left( \frac{8 U}{k_U} + \frac{T}{k_T} + \frac{B}{k_B} - \frac{2 Q}{k_Q} \right) . & 
\label{eq:density_combos}
\eea
where
\bea
C = - \frac{7}{10} \left( \frac{1}{6} Q + \frac{2}{3} T - \frac{1}{3} B - \frac{1}{2} L + H \right),
\eea
obtained through the combination of Eqs.~(\ref{eq:WC_relation}) and~(\ref{eq:x2_conservation_result}). 

Note that we have assumed the rate for EW sphalerons is much slower than all other rates considered thus far and have, thus, not included the EW sphaleron transition terms in computing the densities\footnote{We will take the resulting LH fermion density as input into the EW sphaleron-driven equation for $B+L$ below.}. Consequently, the transport equations should conserve $B+L$. This conservation is manifest for the third generation fermions, as one can see by adding Eqs. (\ref{eq:QTEs2}, \ref{eq:QTEs3},\ref{eq:QTEs5},\ref{eq:QTEs2}) and noting that the transport equation for the RH leptons has a vanishing RHS. For the first and second generation fermions, we note that (a) the transport equation for the first and second generation down-type RH quarks has the same form as Eq.~(\ref{eq:QTEs1}) but with $U\to D$; (b) the equation for the first and second generation LH quark doublets has the same form as Eq.~(\ref{eq:QTEs3}) but with vanishing $\Gamma_H$; (c) the transport equations for the first and second generation LH and RH leptons also have a vanishing RHS. Consequently, $B+L$ is locally conserved for the first and second generations as well in the limit of vanishing EW sphaleron rate.

In Eqs.~(\ref{eq:QTEs1},\ref{eq:QTEs2},\ref{eq:QTEs3},\ref{eq:QTEs5}), the numerical value of the diffusion constant, $D_q$, for all quark states depends on whether SU(3)$_C$ or SU(2)$_C$ is the conserved color symmetry. However, for simplicity, we assume the value $D_q = 6/T$ throughout, obtained in SU(3)$_C$ conserving calculations, while for $D_L$ we take $100/T$ \cite{Joyce:1994zn,Joyce:1994zt} %{\color{blue} reference is redundant. Perhaps the reference for $D_q$ should be removed?}. 
%In the relation $\mathcal{E}_M$ of Eq.~(\ref{eq:density_combos}), the $C$ charge density is actually used as a placeholder for the relation
%\footnote{The difference between these two values is at most an $\mathcal{O}$(1) group factor which does not make a significant difference in our final results.}
%

%We use the closed time path formalism (for a review see, \eg~\cite{}) to calculate all perturbative 
The set of transport coefficients excluding the relaxation term which was already given in Eq.~(\ref{eq:relaxationterm1}) are 
\bea
%& S^{(\mathrm{CPV},\, \mathrm{BLV})} = \ds \frac{ {\rm Im} ( \tilde{y}_1 \tilde{y}_2^* ) (\varphi_1 \dot{\varphi_2} - \varphi_2 \dot{\varphi_1}) }{\pi^2} \mathcal{F}_S & \label{eq:Source} \\
%& \Gamma_M = \ds \frac{3 | \tilde{y}_1 \varphi_1 + \tilde{y}_2 \varphi_2 |^2}{\pi^2 T^3} \mathcal{F}_M & \label{eq:2bodyRelaxation} \\
& \Gamma_H = \ds \frac{36 y_t^2}{T^2} \mathcal{I}_F (m_{t_R}, m_Q, m_C ) + 0.13 \alpha_s T & \\
& \Gamma_C = \ds \frac{144 |\tilde{y}_1|^2 }{T^2} \mathcal{I}_F (m_{b_R}, m_L, m_C ) + 0.52 |\tilde{y_1}| ^2 \alpha_s T . \nonumber \\ & 
\eea

The relaxation rates $\Gamma_H$ and $\Gamma_C$ %{\color{magenta} (1) what are the physical processes that underly both? (2) why are they three body if there are scattering terms? (3) why is there no scattering term for $\Gamma_C$?}  {\color{blue} is this enough detail or should I also explain the scattering processes in more detail?}
depend on the function $\mathcal{I}_F$ \cite{Cirigliano:2006wh} that characterizes the three-body decays, $t_R\to Q+H$ and $b_R \to L +C$ respectively, and a 4-body scattering contribution proportional to $\alpha_s$. 
Note that in regions of mass parameter space where $\mathcal{I}_F$ due to kinematic blocking, the 4-body term remains non-zero. 
Also note that for the sake of simplicity we have restricted ourself to the case where $\tilde{y_1}=\tilde{y_2}$.

%The additional term in $\Gamma_H$ is necessary as $\mathcal{I}_F$ vanishes in certain regions of parameter space due to kinematic threshold effects. In these regions, the additional term, which approximates the effect of the relevant 4-body interactions, will dominate over the 3-body rate and must therefore be included.

Finally, we consider the non-perturbative SU(2)$_C$ strong sphaleron rate $\Gamma_{SS}^{(2)}$.  In Ref.~\cite{Moore:2010jd}, the $N_C$ dependence of the strong sphaleron rate was explored. By following their results, we identify the numerical value of the SU(2)$_C$ strong sphaleron rate to be roughly $\Gamma_{SS}^{(2)} \simeq 9 \alpha_s^4 T$. \\

In the next section, we present our solution of the Boltzmann equations and discuss how this is related to the determination of $Y_B$ in CoBBG.

% Source terms in Boltzmann equations are physically interpreted as reflection and transmission rates for monochromatic quarks and leptons scattering from the bubble wall.

%%%%%%%%%%%%%%%%%%%%%%%%%%%%%%%%%%%%%%%%%%%%%%%%%%%%%%%%%%%%%%%%%%
\subsection{Solving the Quantum Transport Equations and Results}\label{sec:CoBBG3}
%%%%%%%%%%%%%%%%%%%%%%%%%%%%%%%%%%%%%%%%%%%%%%%%%%%%%%%%%%%%%%%%%%

We begin by discussing the parameterization of the set of Boltzmann equations. In principle, the numerical values of all coefficients and source terms in Eq.~(\ref{eq:QTEs}) are parameterized by 6 unknown model parameters: two tree level masses ($m_{H}(T), m_{C} (T)$) 
%{\color{magenta} terrible notation} {\color{blue} is this better?} 
and two complex Yukawa couplings ($\tilde{y}_1, \tilde{y}_2$). However,  only the relative phase, $\delta$,  is physically relevant. Moreover, for simplicity, we assume that both Yukawa couplings have equal magnitudes $\tilde{y} \equiv |\tilde{y}_1| = |\tilde{y}_2|$. Under these assumptions, the prefactors in Eqs.~(\ref{sourceterm}) and~(\ref{eq:relaxationterm}) become
\bea
& {\rm Im} ( \tilde{y}_1 \tilde{y}_2^* ) (\varphi_1 \dot{\varphi_2} - \varphi_2 \dot{\varphi_1}) = \tilde{y}^2 \sin \delta \; \dot{\zeta} \; \varphi_{CB}^2 & \nonumber \\
& {\rm and}& \nonumber \\
& |\tilde{y}_1 \varphi_1 + \tilde{y}_2 \varphi_2|^2 = \tilde{y}^2 \varphi_{CB}^2 \left( 1 + \sin (2 \zeta) \cos \delta \right) &
\eea
respectively, where we remind the reader that $\varphi_{CB}^2  =\varphi_1^2 + \varphi_2^2$ and $\tan \zeta = \varphi_2 / \varphi_1$. \\

The first step in determining $Y_B$ in CoBBG is to solve the Boltzmann equations in Eq.~(\ref{eq:QTEs}). We note that, without a source term for any of the $T$, $Q$, or $U$ densities, there exists a linear combination of their Boltzmann equations for which the right hand side vanishes, \ie 
\bea
v_w (T' + Q' + U') - D_q (T'' + Q'' + U'') = 0 .
\eea
Each of these densities diffuse in the plasma at the same rate which implies that this combination is locally conserved. This means that\begin{equation} 
T + Q + U = 0 \ ,
\end{equation}  
thereby eliminating the need for explicit retention of the transport equation for $U=-(Q+T)$. We then employ the novel methods of Ref.~\cite{White:2015bva} to directly solve the reduced set of five Boltzmann equations analytically without recourse to any assumption about the size of the 3-body rates. 

For each of the densities, $\mathbb{D} = \{T, Q, H, B, L\}$, we assume the boundary conditions $\mathbb{D}(\pm \infty) = 0$. That is, we assume that the electroweak phase transition in which SU(3)$_C$ is restored occurs at a time much larger than the diffusion time scale $\tau _D$. %{\color{magenta} why?} {\color{blue} is this ok?}. 
For simplicity we also approximate the relaxation rate $\Gamma_M$ near the bubble wall as a step function %{\color{magenta} really? What is the motivation and what is the justification for doing so?} 
%{\color{blue} its a bit easier and it produces an error (underestimate) of only $20 \%$ according to my paper. I changed the sentence before and after to discuss this. Can you check this is ok?}
%
\bea
\Gamma_M(z) = \left\{
\begin{tabular}{cc}
0 & $z<0$ (unbroken) \\ \\
$\Gamma_M(10 L_w)$ & $z>0$ (broken)
\end{tabular}
\right.,
\eea
where $L_w$ is the width of the bubble wall and $10 L_w$ is simply a sufficient distance from the wall that $\Gamma_M$ has become constant in the CoB phase. This produces a slight underestimate of the baryon asymmetry, however the error tends to be small \cite{White:2015bva}. In order to determine the numerical values of $\Gamma_M$ and $S^{(\mathrm{CPV},\, \mathrm{BLV})}$ we further require the full spacetime dependence of the CoB vev $\varphi_{CB}$ and its angle $\tan \zeta$ across the bubble wall. A detailed calculation of this dependence requires an involved analysis of the full scalar potential, which we defer to future work. For simplicity, we assume a kink profile \cite{John:1998ip,Akula:2016gpl,Balazs:2016yvi,Akula:2017yfr} %{\color{magenta} this is a common choice; cite references} 
%{\color{blue} I usually cite this reference but you mention plural. Should I cite a bunch from the last 12 months or did you have others specifically in mind?}
%
\bea 
& v_{CB}(z) = \ds \frac{\xi T}{2 \sqrt{2}} \left( 1 + \tanh \left( 2 \alpha \frac{z}{L_w} \right) \right) & \nonumber \\
& \zeta(z) = \ds \frac{\Delta \zeta}{2} \left( 1 + \tanh \left( 2 \alpha \frac{z}{L_w} \right) \right) & 
\eea
where we take $\alpha = \xi = 3/2$. Following detailed calculations in the MSSM \cite{Moreno:1998bq}, we assume a conservative value for $\Delta \zeta \equiv \zeta(T)|_{z \to \infty} - \zeta(T)|_{z \to -\infty}$ which we take to be $\Delta \zeta = 0.01$. Note that the BAU is directly proportional to $\Delta \zeta$ and the presence of scalar singlets can lift the value of $\Delta \zeta$ by an order of magnitude or more \cite{Kozaczuk:2014kva}.

In Fig.~\ref{fig:densityplot}, we present the charge densities for the case of a maximal CP-violating phase, $\sin \delta = 1$, a Yukawa coupling of $\tilde{y} = 0.1$, and tree-level masses of %{\color{magenta} again, terrible notation} 
$m_{C}(T) = 250$ GeV and $m_{H}(T) = 100$ GeV. Moreover, to obtain these results we have used the following phase transition parameters $T = 250$ GeV, $L_w = 10/T$, and $v_w = 0.05$. \\

We now discuss calculation of the total baryon asymmetry in the CoB phase. As previously stated, the baryon asymmetry has two components. A space time varying asymmetry in $B-L$ due to the spontaneous violation of this conserved number within the color broken phase and the usual component that arises from a total left handed number density which biases electroweak sphalerons ahead of the advancing bubble wall producing a net $B+L$ asymmetry. Note that deep within the color broken phase $B+L$ is effectively conserved and this asymmetry will persist into the electroweak phase. 

In Fig. \ref{fig:mvalplot} and Fig. \ref{fig:yvalplot} we show how the baryon asymmetry 
varies as a function of spacetime for various values of the leptoquark tree level mass and coupling respectively. 
The dotted line in these figures are the contribution to the BAU from the electroweak
mechanism whereas the solid line is the space time varying contribution from spontaneous breaking of $B-L$.  The total $B-L$ is zero but there is a non-zero density inside the bubble. Note that the coordinate system is the rest frame of the bubble wall, the $B-L$ density is therefore trapped inside the bubble diluting as the bubble grows.
We normalize the space time variable by the Hubble length at the time of nucleation to highlight that the 
$B-L$ contribution is very small. Generally we find that the space time varying $B-L$ density vanishes at about one trillionth of the Hubble length at the time of nucleation. A significant dilution of this already tiny contribution occurs by the time of recombination.
From these figures we see that one can easily produce the BAU for a large range of parameter space during the color breaking phase
transition. The BAU monotonically decreases with $m_C (T)/T$ but increases with $\tilde{y}$. 
The dependence on $m_C(T)/T$ is gentle indicating a weak dependence on the leptoquark mass. This is explained
by the fact that the leptoquark masses do not enter the functions for the CPV sources, they only appear in the
relaxation term $\Gamma _C$. 
\begin{figure}[t!]
%\centering
\hspace{-.5in}\includegraphics[scale=.6]{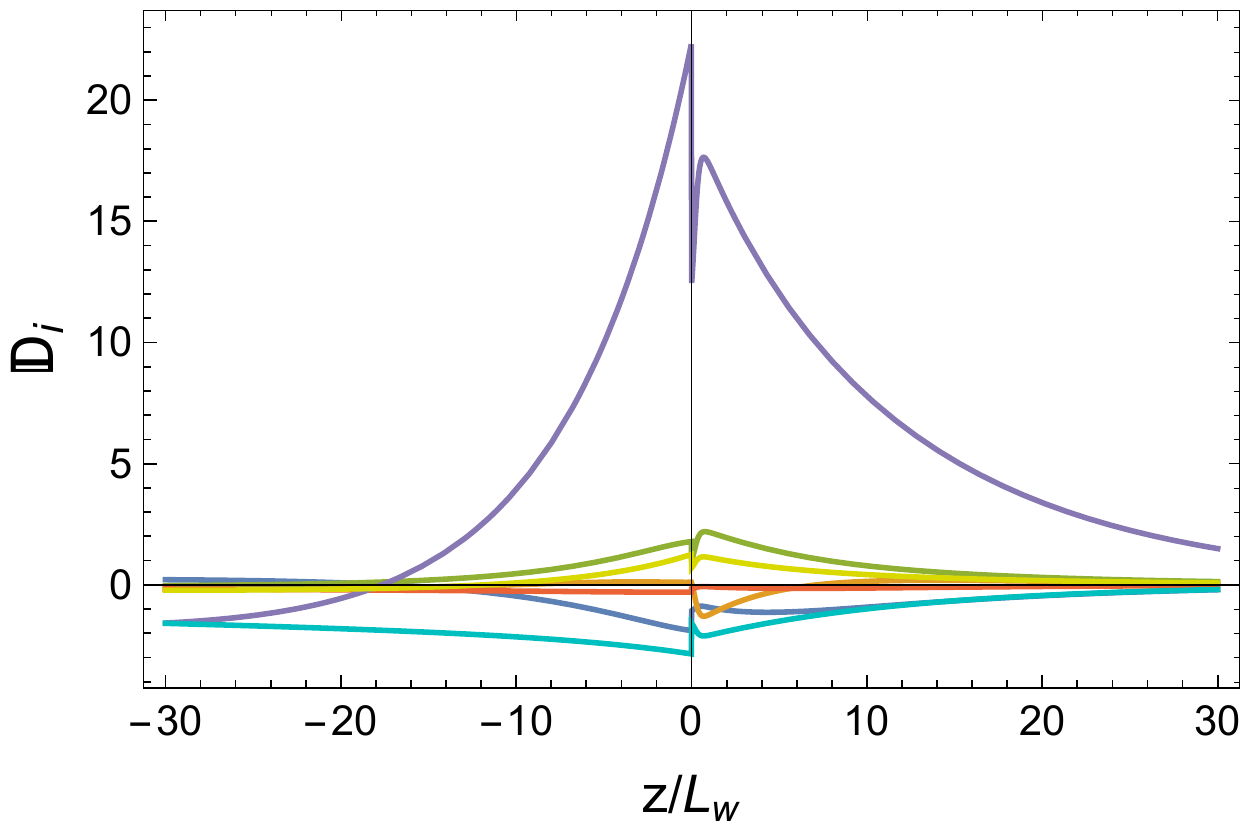}
\caption{Charge densities of all relevant species for $m_{C}(T)=800$ GeV, $T=250$ GeV, $m_{H}(T)=100$ GeV, $\sin \delta =1$, $\tilde{y}=0.1$, $L_w=10/T$ and $v_w=0.05$. Region of positive (negative) $z$ denotes the region of broken (unbroken) SU(3)$_C \times$SU(2)$_L$.}
\label{fig:densityplot}
\end{figure}
%%%%%%%%%%%%%%%%%%%%%%%%%%%%%%%%%%%%%%%%%%%%%%%%%%%%%%%%%%%%%%%%%%

%%%%%%%%%%%%%%%%%%%%%%%%%%%%%%%%%%%%%%%%%%%%%%%%%%%%%%%%%%%%%%%%%%
\begin{figure}[h!]
%\centering
\hspace{-.5in}\includegraphics[scale=.6]{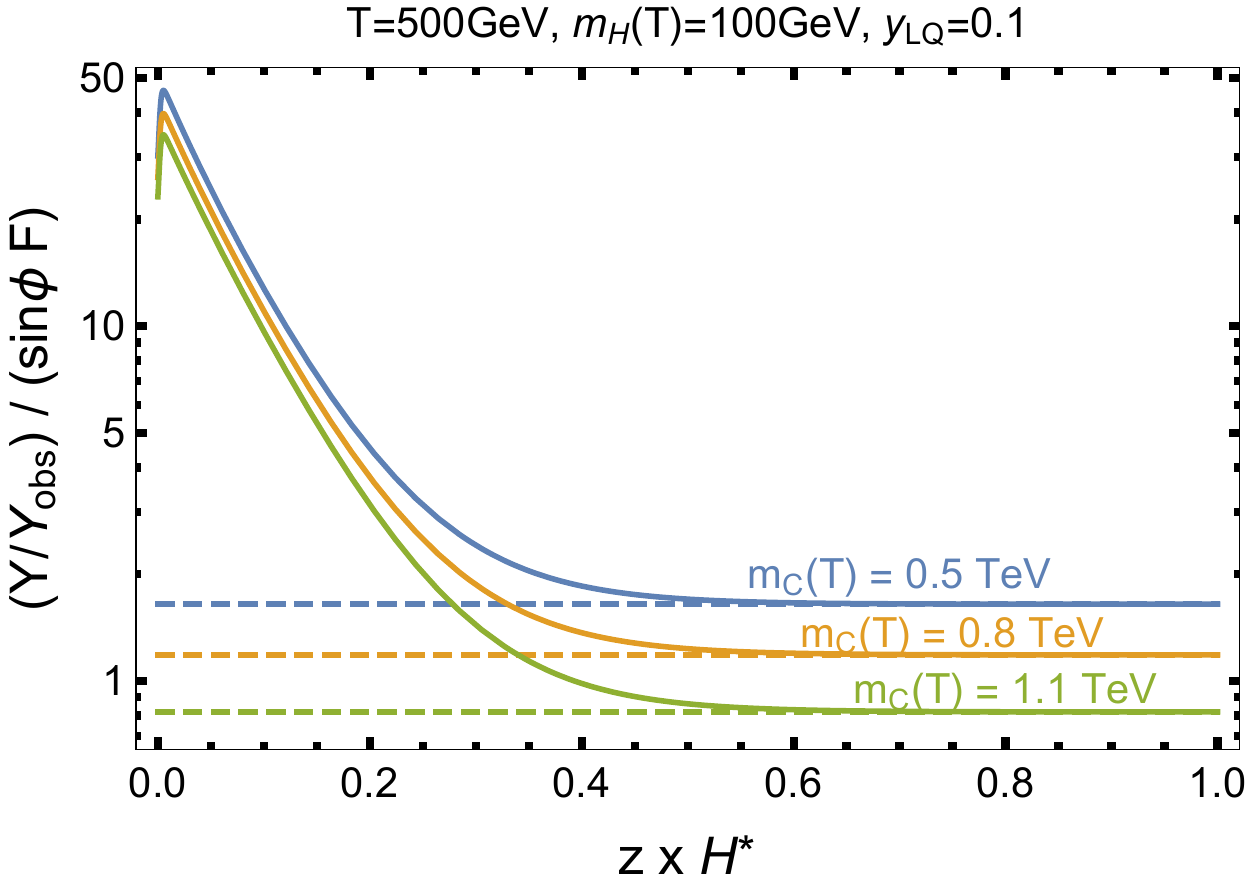}
\caption{Variation in BAU due to $m_{C}(T)$ as a function of the space time variable $z$ normalized by the Hubble length. The BAU has two components: a space time varying component due to the spontaneous violation of baryon asymmetry and a component due to the EWBG mechanism. The space time varying component barely penetrates the bubble wall compared to the Hubble length.}
\label{fig:mvalplot}
\end{figure}
%%%%%%%%%%%%%%%%%%%%%%%%%%%%%%%%%%%%%%%%%%%%%%%%%%%%%%%%%%%%%%%%%%

%%%%%%%%%%%%%%%%%%%%%%%%%%%%%%%%%%%%%%%%%%%%%%%%%%%%%%%%%%%%%%%%%%
\begin{figure}[h!]
\centering
\hspace{-.5in}\includegraphics[scale=.6]{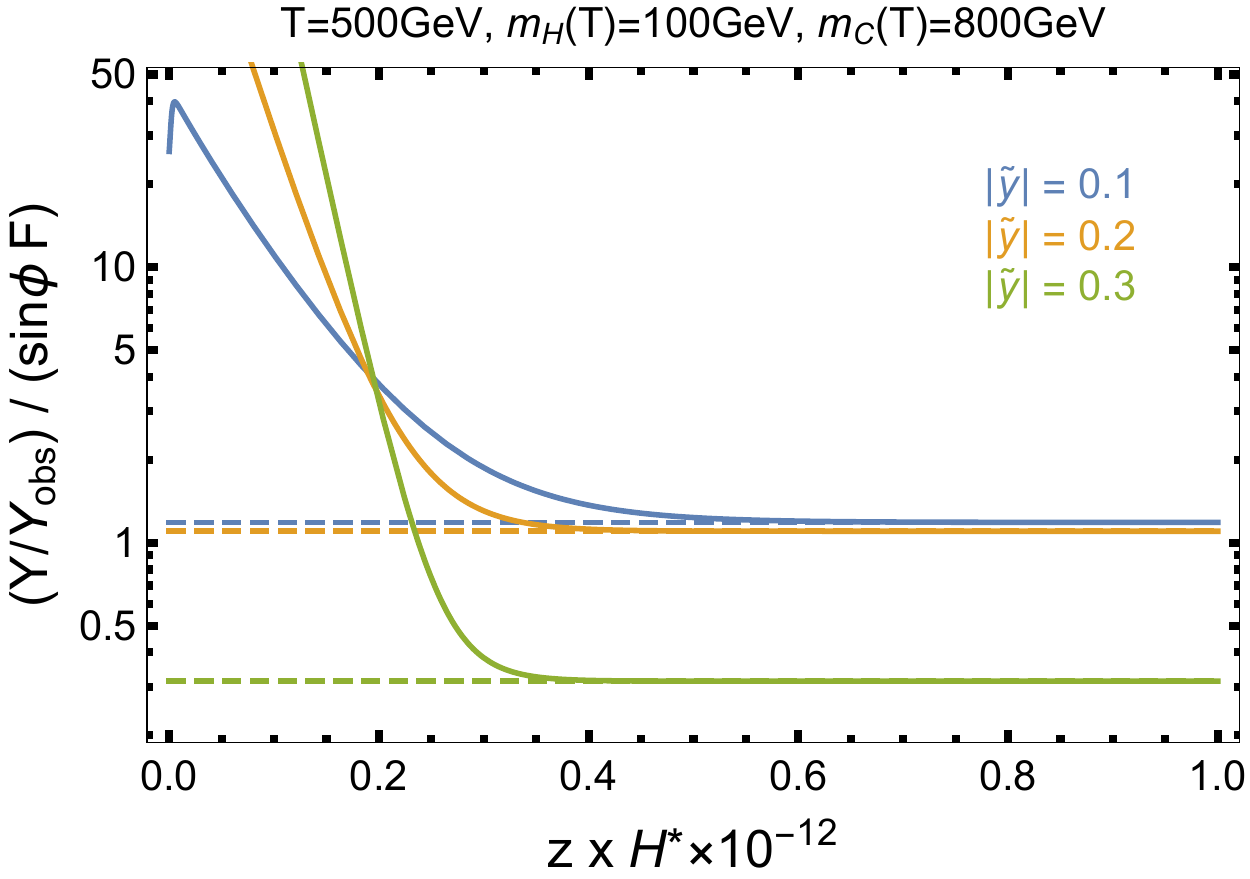}
\caption{Variation in BAU due to $y_{LQ}$ as a function of the space time variable $z$ normalized by the Hubble length. The BAU has two components: a space time varying component due to the spontaneous violation of baryon asymmetry and a component due to the EWBG mechanism. The space time varying component barely penetrates the bubble wall compared to the Hubble length. }
\label{fig:yvalplot}
\end{figure}
\section{Phenomenology}\label{sec:Pheno}
%%%%%%%%%%%%%%%%%%%%%%%%%%%%%%%%%%%%%%%%%%%%%%%%%%%%%%%%%%%%%%%%%%
\subsection{LHC constraints}
At $T=0$, the colored scalars, $C_1$ and $C_2$, are produced through their strong interactions at the LHC. Under the assumption given in Eq.~(\ref{eq:yuk_texture}), the scalar decay modes are $\chi^{2/3} \to b_R \tau_L$ and $\chi^{-1/3} \to b_R \nu_{\tau L}$ with unit branching ratios. The CMS collaboration has recently placed limits on scalar leptoquarks which dominantly decay into these modes by studying their pair production. The dominant pair production mechanisms at the LHC for these colored scalars are through gluon-gluon fusion and quark-antiquark annihilation, for which the cross sections depend only on the scalar mass. At  $\sqrt{s}$=8 TeV, limits have been derived on colored scalars decaying to $b_R \tau_L$~\cite{Sirunyan:2017yrk} %and $b_R \overline{\nu}_\tau$~\cite{Chatrchyan:2012st}
using an integrated luminosity of 12.9 fb$^{-1}$. %In each case, 
A unit branching ratio was assumed and upper limits on the production cross sections were set at the 95\% C.L., yielding the bound of $m_{C_{2/3}} \geq $850 GeV. %and $m_{C_{-1/3}} \geq $440 GeV. 
Limits on leptoquarks decaying in the $b_R \overline{\nu}_\tau$ mode derived by the ATLAS collaboration are $m_{C_{-1/3}} \geq $640 GeV~\cite{Aad:2015caa}. 

Aside from direct searches, the colored scalars of CoBBG can also be searched for indirectly by examining their effects on the rates for production and decay of the SM Higgs. At the 1-loop level, their SU(3)$_C$ charges enable them to interfere with top quark loops in gluon-gluon fusion production of the SM Higgs. As well, their U(1)$_{EM}$ charges enable them to interfere with both top quark and $W^\pm$ loops in Higgs-to-diphoton decay. The modifications of these rates are best expressed as ratios with the SM-valued rates, $R_{\gamma \gamma} (R_{gg}) \equiv \Gamma_{\gamma \gamma } / \Gamma_{\gamma \gamma}^{SM} (\sigma_{gg} / \sigma_{gg}^{SM})$. At leading, non-trivial order, one has
\begin{eqnarray}
& \ds R_{\gamma \gamma} = \frac{ \left| F_1 ( \tau_W ) + \frac{4}{3} F_{1/2} (\tau_t)  + N_c \sum_{i} Q_{EM}^2 \xi_{C_i} F_0 (\tau_{C_i}) \right|^2 }{ \left| F_1 ( \tau_W ) + \frac{4}{3} F_{1/2} (\tau_t) \right|^2 } & \nonumber \\ \nonumber \\
& \ds R_{gg} = \frac{ \left| F_{1/2} (\tau_t) +  \sum_{i} \xi_{C_i} F_0 (\tau_{C_i}) \right|^2 }{ \left|  F_{1/2} (\tau_t) \right|^2 } ,
\end{eqnarray}
where we sum over the contributions of each colored scalar. Here, we have defined $\tau_i$=$4 m_i^2 / m_h^2$, 
\begin{equation}
\xi_{C_i}=2 \frac{\lambda_{HC_i}}{g_1} \frac{M_W^2}{m_{C_i}^2}\ \ \ ,
\end{equation}
 $Q_{EM}$ is the electric charge of the scalar $C_i$, and all loop functions are defined in Ref.~\cite{Gunion:1989we}.  The parameters $\lambda_{HC_i}$ are the couplings associated with the Higgs portal operator $H^\dagger H C_i^\dagger C_i$. While they do not directly enter the transport computation, they are nevertheless important for the phase transition dynamics. 

Using these ratios, we construct the set of signal rates $\mu_{XX}$ associated with Higgs measurements, relative to pure SM-Higgs expectations, \ie 
\begin{eqnarray}
\mu_{XX} = \frac{ \sigma \cdot \text{BR}  }{  \sigma^{SM} \cdot \text{BR}^{SM} } .
\end{eqnarray} 
Each signal rate is a function of the Higgs portal couplings $\lambda_{HC_i}$ and scalar masses $m_{C_i}$ and, for simplicity, we assume that all scalars are degenerate in mass and share the same $\lambda_{HC}$. We then impose constraints on these parameters by performing a global $\chi^2$ fit to the current Higgs data\footnote{aside from \cite{CMS:2017tkf,ATLAS:2017ovn,Aaboud:2017xsd} which uses 13 TeV data, the most up to date signal strengths are taken from 7 and 8 TeV data in Ref. \cite{Khachatryan:2016vau}. We use the 13 TeV signal strengths and uncertainties unless it is unavailable.} using 
\begin{eqnarray}
\chi^2 ( \lambda_{HC}, m_C ) = \sum_X  \left( \frac{ \mu_{XX}^{obs} - \mu_{XX} }{ \Delta \mu_{XX}^{obs}  } \right)^2 ,
\end{eqnarray}
where $\mu_i^{obs}$ ($\Delta \mu_i^{obs}$) are the (uncertainties in the) observed signal rates. The resulting 95\% C.L. limit on the parameters, shown in Fig.~\ref{fig:HiggsFit}, implies that, for scalar masses above the current direct search limits ($m_{C} \gtrsim$500 GeV), a wide range of $\lambda_{HC}$ is open. We also include future projected limits expected from the HL-LHC~\cite{ATL-PHYS-PUB-2013-014,CMS:2013xfa,HLLHC1,DeAlmeidaDias:2055285}, represented by the solid and dashed black contours in Fig.~\ref{fig:HiggsFit}. \\

%%%%%%%%%%%%%%%%%%%%%%%%%%%%%%%%%%%%%%%%%%%%%%%%%%%%%%%%%%%%%%%%%%
\begin{figure}[t!]
%\centering
\hspace{-.5in}\includegraphics[scale=.6]{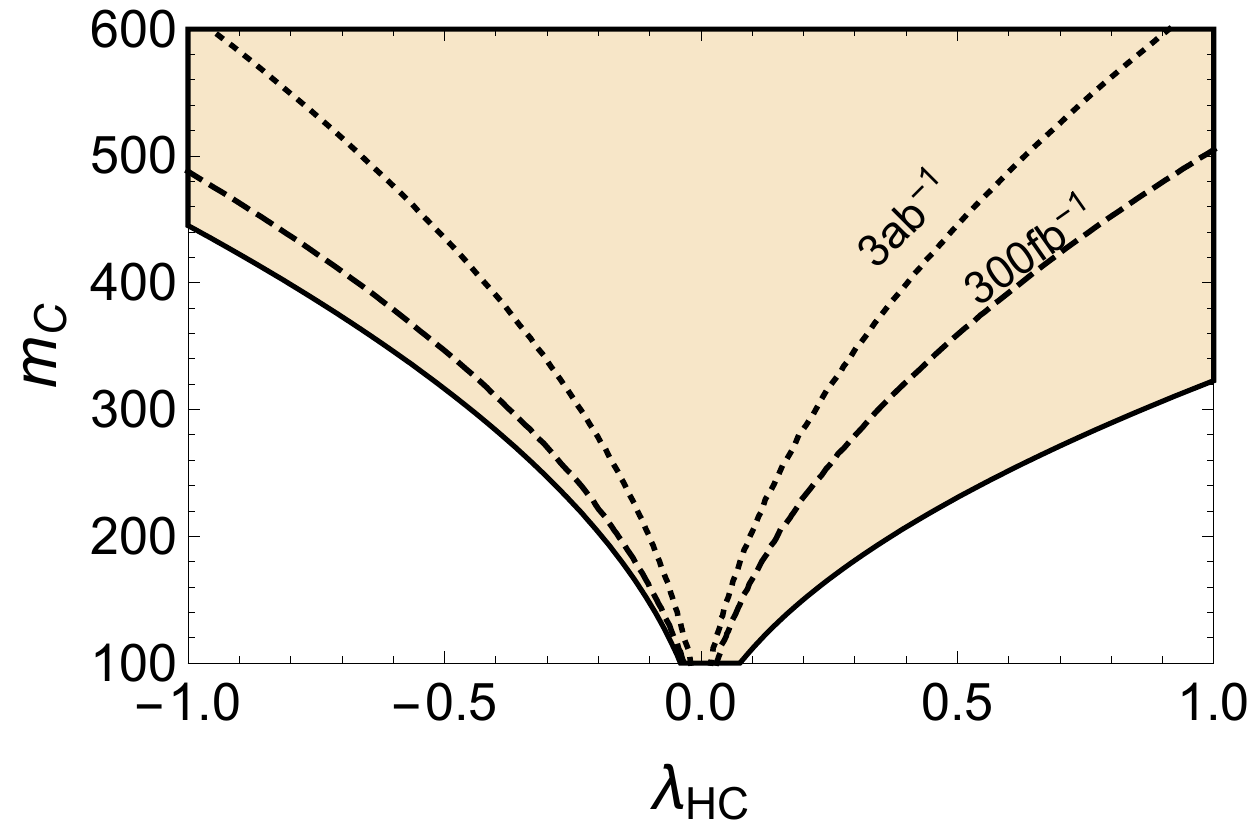}
\caption{Shaded region represents the allowed ($m_C$,$\lambda_{HC}$) parameter space from current LHC Higgs measurements at the 95\% C.L. The dashed (dotted) line represents the 95\% C.L. projected sensitivity to this parameter space at the 300fb$^{-1}$ (3ab$^{-1}$) high luminosity LHC. }
\label{fig:HiggsFit}
\end{figure}
%%%%%%%%%%%%%%%%%%%%%%%%%%%%%%%%%%%%%%%%%%%%%%%%%%%%%%%%%%%%%%%%%%

\subsection{Electric Dipole Moments}\label{EDMs}

%%%%%%%%%%%%%%%%%%%%%%%%%%%%%%%%%%%%%%%%%%%%%%%%%%%%%%%%%%%%%%%%%%%%%%%%%%%%%%%%%%%%%%%%%%%%%%%%%%%%%%%%%%%%%%%%%%%
%%%%%%%%%%%%%%%%%%%%%%%%%%%%%%%%%%%%%%%%%%%%%%%%%%%%%%%%%%%%%%%%%%%%%%%%%%%%%%%%%%%%%%%%%%%%%%%%%%%%%%%%%%%%%%%%%%%

%
%\begin{figure}[t!]
%\centering
%\includegraphics[scale=.65]{figs/HiggsFit.pdf}
%\caption{\textit{Current 95\% C.L. limits on the ($m_C$,$\lambda_{HC}$) parameter space from current Higgs measurements at the LHC. Shaded region is allowed. For masses $m_C \gtrsim$ 500 GeV, a large range of $\lambda_{HC}$ values are available.}}
%\label{fig:Higgs}
%\end{figure}
%
Searches for permanent electric dipole moments (EDMs) provide constraints on the CP violating phases necessary for producing a baryon asymmetry (see, {\em e.g.}, Ref.~\cite{Morrissey:2012db} and references therein. For other recent EDM reviews, see \cite{Pospelov:2005pr,Engel:2013lsa,musolfedm,musolfunpub} %{\color{magenta} include the review papers by Engel, MJRM and van Kolck and Pospelov \& Ritz}). 
Here, we consider EDM constraints on  the CP violating phases present in the leptoquark couplings, $y _i$. We find that improvements in experimental sensitivity by many orders of magnitude would be needed to probe the full parameter space of the specific CoB scenario discussed here. 

%The conclusion we will argue is that current searches for permanent EDMs do not constrain the CP violating phases in our model and in fact will need to increase in sensitivity by $2-3$ orders in magnitude in order to probe the parameter space. A discussion of how much this is generically true within the paradigm of color breaking baryogenesis compared to it merely being a feature of the particular model we have chosen we leave to future work.

 We work in the effective field theory frame work where weak scale particles, $t$, $W^\pm$, $Z$, $H$, and $C_{1,2}$ are considered heavy and integrated out. The effective Lagrangian that results from this is a sum of fermion EDMs, chromo-EDMs and the three-gluon Weinberg operator~\cite{Weinberg:1989dx}
\begin{eqnarray}
\mathscr{L}_{CPV} &=& - \frac{i}{2} d_f \overline{f} \sigma^{\mu \nu} \gamma^5 f F_{\mu \nu} - \frac{i}{2} \tilde{d}_q g_s \overline{q}_i \sigma^{\mu \nu} \gamma^5 (T^a)_{ij} q_j G_{\mu \nu}^a \nonumber \\
&+& g_s\frac{C_W}{\Lambda^2} f^{abc} G _{\mu \nu} ^a \tilde{G} ^{b \nu \lambda} G^{c \mu} _\lambda + h.c.
\label{CPVLag}
\end{eqnarray}
Here, $F_{\mu \nu}$ ($G_{\mu \nu}^a$) is the photon (gluon) field strength, $\tilde{G}^a_{\mu \nu} \equiv \frac{1}{2} \epsilon_{\mu \nu \alpha \beta} G^{a \alpha \beta}$ is the dual field strength (with $\epsilon_{0123} = +1$), and $T^a$ and $f^{abc}$ are the full SU(3)$_C$ generators and structure constants, respectively. Finally $\Lambda$ is the BSM scale that has been explicitly factored out, whereas the coefficients of the dipole operators retain dimensions of one inverse power of the mass. We will assume that the QCD $\theta$ term, arising at dimension-four in the SM, is removed by the Peccei-Quinn mechanism~\cite{Peccei:1977hh}. Moreover, we do not consider CP-odd four-fermion interactions (generated by tree-level $C_{1,2}$ exchange) as all couplings to first and second generation fermions are suppressed.

Elementary fermion EDMs and CEDMs originate first at the 3-loop level from the three loop, Barr-Zee type graphs shown in Fig.~\ref{fig:EDM} [panel a]. This loop suppression arises from the need for a different phase in the Yukawa vertices in the fermion loop. Such a diagram requires mixing of $C_{1,2}$ mixing through Higgs portal interactions\footnote{Note that inclusion of unsuppressed first and second generation leptoquark interactions one give rise to one-loop elementary fermion (chromo-) EDMs}. Naive dimensional analysis  yields for the electron EDM
\begin{eqnarray}
d_e &\simeq & e \frac{\alpha_{EM}}{4 \pi} \frac{\text{Im}(y_1 y_2)}{(4 \pi)^4} \frac{m_e m_b^2}{m_C^4} \nonumber \\
&\sim & 5 \times 10^{-36} \; \text{Im}(y_1 y_2) \left( \frac{\text{TeV}}{m_C} \right)^4 e \cdot \text{cm} .
\end{eqnarray}
The scaling with $m_C^{-4}$ is not surprising, since the $C_1$-$C_2$ mixing an insertion of $\langle H^\dag H\rangle$ associated with mixing in the scalar potential. For $m_C = 500$ GeV this gives $d_e \sim 10^{-34} \; \text{Im}(y_1 y_2)~e \cdot \text{cm}$, far below even the recent ACME bound $|d_e| < 8.7 \times 10^{-29} \; e \cdot \text{cm}$~\cite{Baron:2013eja}, leaving the CP-odd phases unconstrained. Accordingly, we neglect all fermion (chromo-)EDMs.

The neutron EDM receives contributions from the quark EDM and chromo-EDM operators as well as the Weinberg  three-gluon operator. For the light quarks, the (chromo-)EDMs will be enhanced compared to $d_e$ by $m_q/m_e\sim 10$, where $m_q$ is the light quark mass. The resulting contribution will, nevertheless, be far too small to be experimentally relevant. 

%{\color{magenta} the entire following analysis and discussion of the Weinberg three gluon operator needs to be re-done in light of more recent work as well as the mercury EDM: (a) the review paper with Engel, MRM and van Kolck gives the sensitivity of $d_n$ to the Weinberg operator, including the theoretical uncertainty; (b) the running has been computed recently by Dekens and de Vries as well as by Hisano; (c) the contribution to mercury is dominated by the isocalar P- and T-odd pion-nucleon coupling, also discussed in the review paper}

The neutron EDM, $d_N$, and the isoscalar $P$ and $T$ odd pion-nucleon coupling $\bar{g}_\pi ^0$ are both sensitive to the Weinberg operator. Following Ref.~\cite{Engel:2013lsa} one can relate both $d_N$ and $\bar{g}_\pi ^0$ to the Wilson coefficient of the Weinberg operator
\begin{eqnarray}
d_N &=&   \frac{v^2}{m_C^2}  Im[C_W] \beta _{\tilde{G}} \label{EDMestimate} \\
{\bar g}_\pi ^0&=& \frac{v^2}{m_C^2}   Im[C_W] \gamma _{\tilde{G}} \label{ISOestimate}
\end{eqnarray}
where
\begin{equation}
 \beta _{\tilde{G}} = [2-40] \times 10^{-7} e \cdot fm, \quad \gamma  _{\tilde{G}}= [1-10] \times 10^{-6}\ .
\end{equation}
%where the uncertainty in this result is approximately a factor of 2 and $d_G$ is to be evaluated at the hadronic scale, that is, $\mu_H = 1$ GeV.
Based on the diagram in Fig.~\ref{fig:EDM} [panel b], we estimate $C_W$ at the weak scale, obtaining %{\color{magenta} I think the following should have an additional power of $(v/m_C)^2$ because the $C_1$-$C_2$ mixing requires the same $\langle H^\dag H\rangle$ insertion that appears in the Barr-Zee graph}
%
%\begin{eqnarray}
%d_G \simeq \frac{g_s^3}{(4 \pi)^4} \frac{\text{Im}(y_1 y_2)}{ m_C^2} f \left( \frac{m_C^2}{m_b^2} \right)  \ .
%\label{EDMestimate}
%\end{eqnarray}
%
\begin{equation}
 C_W=   \frac{ g_s^2}{(4 \pi)^4}{\rm Im}(y_1y_2)f\left( \frac{m_C^2}{m_b^2} \right) \ .
\end{equation}
Here $f$ is a 2-loop function which we identify with that calculated in Ref.~\cite{Dicus:1990ab}. We emphasize that, since the internal scalar lines themselves have SU(3)$_C$ charge and can also emit gluons, the true loop function inevitably differs from that of Ref.~\cite{Dicus:1990ab}. However, we expect such contributions to be suppressed relative to Eqs.~(\ref{EDMestimate}) and (\ref{ISOestimate}) due to their explicit momentum dependence so we persist for now with the above estimate.

Following Ref.~\cite{Degrassi:2005zd,Hisano:2012cc,Dekens:2013zca}, the running of the Weinberg operator coefficient from the weak scale to the hadronic scale is given by
\begin{equation}
 C_W (M_{\rm QCD})= \left(\frac{\alpha _s (M_W)}{\alpha _s(M_{\rm QCD})} \right) ^{\gamma _G/(2\beta _0)} C_W(M_W) \ . 
\end{equation}
with anomalous dimension $\gamma_G = N_c + 2 n_f + \beta_0$, $\beta_0 = 11 - 2/3 n_f$, and $n_f \equiv$ the number of active quark flavors. As heavy quark flavors are integrated out at their respective masses, threshold effects arise~\cite{Braaten:1990gq}, inducing a shift in $C_W$ proportional to the corresponding CEDM. In particular, such a shift occurs at the $b$-quark mass threshold and can lead to significant effects if $\tilde{d}_b$ can be generated at the 1-loop level. However, as $C_{1,2}$ couple only to $b_R$, generation of $\tilde{d}_b$ still arises from the Bar-Zee graphs in Fig.~\ref{fig:EDM} [panel a], rendering the resulting shift completely negligible. 

The resulting estimate for the neutron EDM and $\bar{g}_\pi ^0$ are then
\begin{eqnarray}
d_N &\approx& [3-60]\times10^{-25}  \frac{v^2}{m_C^2} {\rm Im}(y_1y_2)f\left( \frac{m_C^2}{m_b^2} \right)  e \cdot cm \nonumber \\
\bar{g}_\pi ^0 &\approx& [1.5-15] \times 10^{-11}  \frac{v^2}{m_C^2} {\rm Im}(y_1y_2)f\left( \frac{m_C^2}{m_b^2} \right) 
\end{eqnarray}
which, for $m_C = 500$ GeV, gives $d_N \simeq 10^{-28} \; \text{Im}(y_1 y_2)$. The current upper limit on the neutron EDM is set at the 90\% C.L. as $|d_n| < [2.9-3.0] \times 10^{-26} \; e \cdot cm$~\cite{Baker:2006ts,Afach:2015sja}, implying that next-generation neutron EDM experiments require improvements of $\mathcal{O}$($10^2$-$10^3$) to directly probe the CP violation responsible for baryon production during the CoB phase transition. The current upper limit on the isoscalar coupling is $|\bar{g}_\pi ^0| < 3.8 \times 10^{-12} $ \cite{musolfunpub} and we similarly find we are at least two orders of magnitude below this bound for a leptoquark mass of $500$ GeV. So contributions to EDMs are indeed constrained within our model. However, we make the following two caveats to our analysis

\begin{itemize}
\item[$\bullet$] Under our current assumption of a CP-conserving potential, the operators responsible for $C_{1,2}$ mixing do not themselves contribute a phase and thus precision measurements of EDMs directly constrain the phase responsible for baryon production during the CoB transition. If this assumption is relaxed, the connection between the CP violation responsible for baryon production and that appearing in EDMs becomes less clear.
\item[$\bullet$] The values of $\beta_{\tilde{G}}$ and $\gamma _{\tilde{G}}$ are quite uncertain and span an order of magnitude.
%{\color{magenta} Change this to reflect the range in the Engel, MJRM, van Kolck review and simply note that the contribution is quite uncertain} The Weinberg coefficient is suppressed at low scales, i.e., $d_G(\mu_H) \simeq 0.4 d_G(\mu_W)$ from Eq.~(\ref{dGRunning}). This differs from Ref.~\cite{Ellis:2008zy} which claimed an enhancement of $\sim$8.5, amounting to an $\mathcal{O}(20)$ overestimate of its importance for $d_n$.
\end{itemize}
We leave a thorough calculation of each EDM as well as consideration of these issues to a future project.
\begin{figure}[t!]
\centering
\includegraphics[scale=.325]{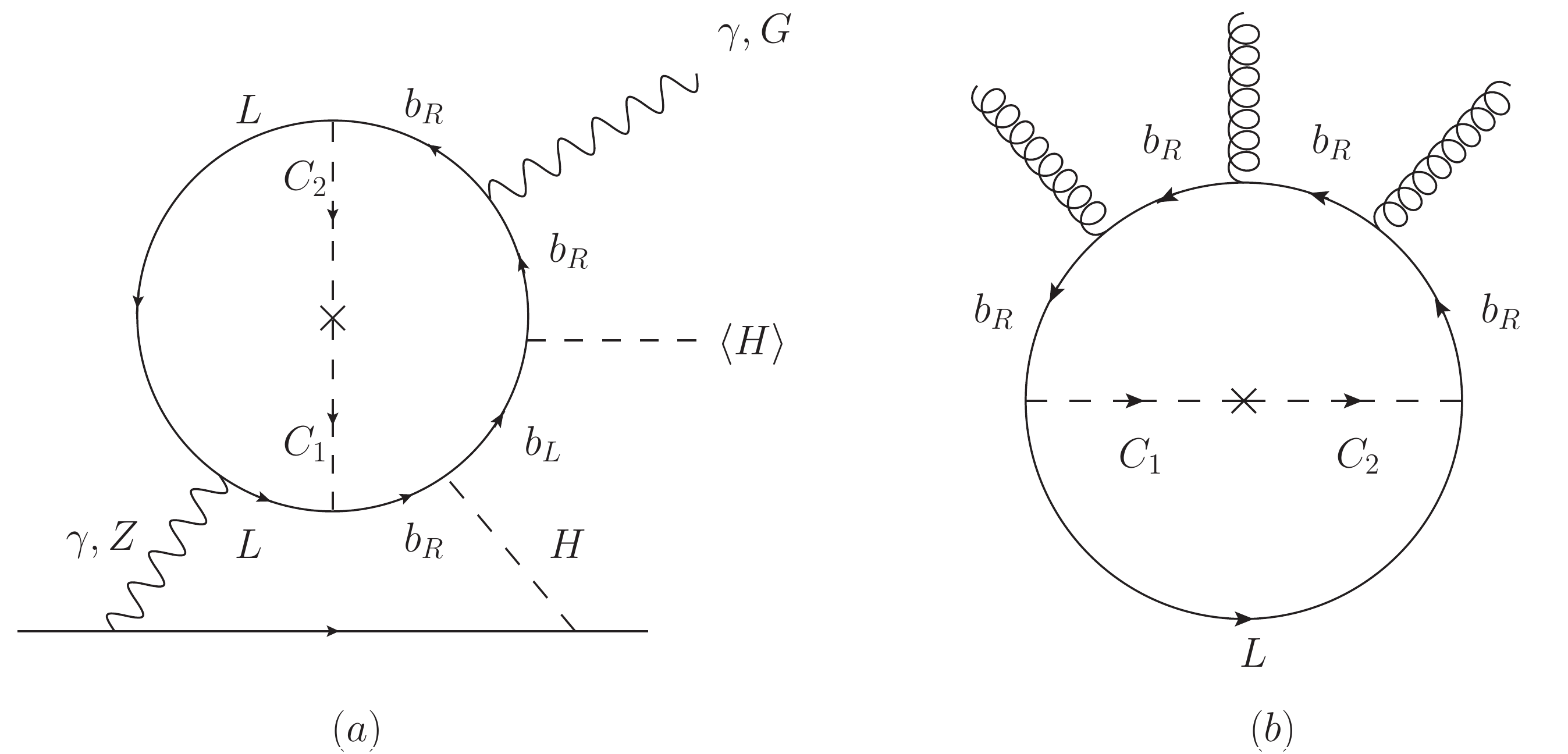}
\caption{Panel (a): Fermion EDMs and chromo-EDMs arise at the 3-loop level, precluding any resulting constructive bounds on the parameter space from these sources. Panel (b): The neutron EDM is sensitive to CP violation in CoBBG through the Weinberg operator. Next generation experiments searching for neutron EDMs require an improvement of roughly $\mathcal{O}(10^3)$ in order to test the CoBBG scenario.}
\label{fig:EDM}
\end{figure}
%

%%%%%%%%%%%%%%%%%%%%%%%%%%%%%%%%%%%%%%%%%%%%%%%%%%%%%%%%%%%%%%%%%%

 %%%%%%%%%%%%%%%%%%%%%%%%%%%%%%%%%%%%%%%%%%%%%%%%%%%%%%%%%%%%%%%%%%
\section{Conclusions}\label{Conclusions}
EWBG links the generation of the cosmic baryon asymmetry to electroweak symmetry breaking in the early universe. In contrast to other theoretically well-motivated scenarios, it is one of the most testable, since it involves new weak scale physics. Not surprisingly, null results for permanent EDMs as well as new particle searches at the LHC tightly constrain EWBG models. 

In this work we have relaxed the assumption that today's symmetries have always been symmetries of nature throughout our cosmic history \cite{Weinberg:1974hy,Mohapatra:1979qt,Mohapatra:1979vr,Langacker:1980kd,Hammerschmitt:1994fn,Dvali:1995cj,Dvali:1996zr,Cline:1999wi,Patel:2012pi,Patel:2013zla,Blinov:2015sna}. We specifically examine the possibility that SU(3)$_C$ was broken for a period and then subsequently restored \cite{Patel:2013zla}. This framework of CoB represents a new EWBG paradigm. We have presented an implementation of this framework that successfully reproduces the BAU without significant fine-tuning while evading present experimental constraints. The framework is still testable because the leptoquark couplings cannot be arbitrarily small, nor can their masses be arbitrarily heavy.

 In CoB, the BAU is generated during an intermediate color breaking phase transition. We consider the case where color-breaking fields couple to SM fermions so as to avoid stable colored relics. Furthermore, in our implementation the interaction between the color breaking fields and the standard model fermions conserve $B-L$. As such, the spontaneous breaking and restoration of SU(3)$_C$ is associated with spontaneous breaking and restoration of $B-L$. However, during the color-breaking transition a $B+L$ asymmetry is generated through the electroweak mechanism which persists even when $B-L$ is restored.  % The spontaneous breaking of $B-L$ leaves a very small relic asymmetry that is associated with relic charge asymmetry that is well below observational bounds
The contribution from the spontaneous violation of $B-L$ is negligible as any such contribution is quickly relaxed away from the bubble wall. 

We conclude by noting that our particular implementation of CoB was a proof of concept. There are other possible implementations of CoB and to truly test the viability of any particular model one would need to simultaneously examine the phase transition and the transport dynamics. We leave such an examination to future work.

%%%%%%%%%%%%%%%%%%%%%%%%%%%%%%%%%%%%%%%%%%%%%%%%%%%%%%%%%%%%%%%%%%

\begin{acknowledgments}
We thank Lorenzo Sorbo for helpful discussions. 
GW would like to acknowledge that his contribution to this work was partly funded by both the American Australian Association via the Keith Murdoch fellowship as well the Australian Postgraduate Award (APA). MJRM and PW were supported in part under U.S. Department of Energy contract DE-SC0011095.
\end{acknowledgments}

\bibliographystyle{h-physrev3.bst}
\bibliography{cobRefs}

\section*{Appendix}

Here, we provide a table of charge values associated with all symmetries in the symmetric, CoB, and electroweak phases for all particle species.

\begin{table*}[t!]
\begin{tabular}{| c | c | c | c | c | c |c|}
\hline
State & $\mathcal{Q}_{T^3}$ & $\mathcal{Q}_{T^8}$ & $\mathcal{Q}_{\tau^3}$ & $\mathcal{Q}_{Y}$ & $\mathcal{Q}_{X_1} = \mathcal{Q}_{T^8} - \frac{2}{\sqrt{3}} \mathcal{Q}_{\tau^3}$ & $\mathcal{Q}_{X_2} = \mathcal{Q}_{\tau^3} + 3 \mathcal{Q}_Y$ \\
\hline
$\{ u_{1 L}, u_{2 L}, u_{3 L} \}$ & $\{ \frac{1}{2}, -\frac{1}{2}, 0 \}$ & $\{ \frac{1}{2 \sqrt{3}}, \frac{1}{2 \sqrt{3}}, -\frac{1}{\sqrt{3}} \}$ & $\frac{1}{2}$ & $\frac{1}{6}$ & $\{ -\frac{1}{2 \sqrt{3}}, -\frac{1}{2 \sqrt{3}}, -\frac{2}{\sqrt{3}} \}$ & 1 \\
\hline
$\{ d_{1 L}, d_{2 L}, d_{3 L} \}$ & $\{ \frac{1}{2}, -\frac{1}{2}, 0 \}$ & $\{ \frac{1}{2 \sqrt{3}}, \frac{1}{2 \sqrt{3}}, -\frac{1}{\sqrt{3}} \}$ & $-\frac{1}{2}$ & $\frac{1}{6}$ & $\{ \frac{\sqrt{3}}{2}, -\frac{ \sqrt{3}}{2}, 0 \}$ & 0 \\
\hline
$\{ u_{1 R}, u_{2 R}, u_{3 R} \}$ & $\{ \frac{1}{2}, -\frac{1}{2}, 0 \}$ & $\{ \frac{1}{2 \sqrt{3}}, \frac{1}{2 \sqrt{3}}, -\frac{1}{\sqrt{3}} \}$ & $0$ & $\frac{2}{3}$ & $\{ \frac{1}{2 \sqrt{3}}, \frac{1}{2 \sqrt{3}}, -\frac{1}{\sqrt{3}} \}$ & 2 \\
\hline
$\{ d_{1 R}, d_{2 R}, d_{3 R} \}$ & $\{ \frac{1}{2}, -\frac{1}{2}, 0 \}$ & $\{ \frac{1}{2 \sqrt{3}}, \frac{1}{2 \sqrt{3}}, -\frac{1}{\sqrt{3}} \}$ & $0$ & $-\frac{1}{3}$ & $\{ \frac{1}{2 \sqrt{3}}, \frac{1}{2 \sqrt{3}}, -\frac{1}{\sqrt{3}} \}$ & -1 \\
\hline
$e_L$ & $0$ & $0$ & $-\frac{1}{2}$ & $-\frac{1}{2}$ & $\frac{1}{\sqrt{3}}$ & -2 \\
\hline
$\nu_L$ & $0$ & $0$ & $\frac{1}{2}$ & $-\frac{1}{2}$ & $-\frac{1}{\sqrt{3}}$ & -1 \\
\hline
$H^+$ & $0$ & $0$ & $\frac{1}{2}$ & $\frac{1}{2}$ & $-\frac{1}{\sqrt{3}}$ & 2 \\
\hline
$H^0$ & $0$ & $0$ & $\frac{1}{2}$ & $\frac{1}{2}$ & $\frac{1}{\sqrt{3}}$ & 1 \\
\hline
$W^\pm_\mu$ & 0 & 0 & $\pm 1$ & 0 & $ \mp \frac{2}{\sqrt{3}}$ & $\pm 1$ \\
\hline
$\{ \chi_1^{2/3}, \chi_2^{2/3}, \chi_3^{2/3} \}$ & $\{ \frac{1}{2}, -\frac{1}{2}, 0 \}$ & $\{ \frac{1}{2 \sqrt{3}}, \frac{1}{2 \sqrt{3}}, -\frac{1}{\sqrt{3}} \}$ & $\frac{1}{2}$ & $\frac{1}{6}$ & $\{ -\frac{1}{2 \sqrt{3}}, -\frac{1}{2 \sqrt{3}}, -\frac{2}{\sqrt{3}} \}$ & 1 \\
\hline
$\{ \chi_1^{-1/3}, \chi_2^{-1/3}, \chi_3^{-1/3} \}$ & $\{ \frac{1}{2}, -\frac{1}{2}, 0 \}$ & $\{ \frac{1}{2 \sqrt{3}}, \frac{1}{2 \sqrt{3}}, -\frac{1}{\sqrt{3}} \}$ & $-\frac{1}{2}$ & $\frac{1}{6}$ & $\{ \frac{\sqrt{3}}{2}, -\frac{ \sqrt{3}}{2}, 0 \}$ & 0 \\
\hline
$\{ G^{\pm,12}_\mu, G^{\pm,45}_\mu, G^{\pm,67}_\mu \}$ & $\{ \pm 1, \pm 1/2, \mp 1/2 \}$ & $\{ 0, \pm \frac{\sqrt{3}}{2}, \pm \frac{\sqrt{3}}{2} \}$ & 0 & 0 & $\{ 0, \pm \frac{\sqrt{3}}{2}, \pm \frac{\sqrt{3}}{2} \}$ & 0 \\
\hline
\end{tabular}
\caption{\it Table of charges for all species in the plasma. Singular values for a given charge imply that all colors have the same charge.}
\label{table:charge_table}
\end{table*}

\end{document}